\documentclass[10pt,journal]{IEEEtran}
\ifCLASSINFOpdf

\else
\fi
\usepackage{graphicx,amsmath,amssymb,cite}
\usepackage{color}
\usepackage{amsfonts}
\usepackage{amsmath}
\usepackage{graphicx}
\usepackage{amssymb}
\usepackage{stfloats}
\usepackage{subfig}
\usepackage{mathrsfs}
\usepackage{algorithm}
\usepackage{algorithmicx}
\usepackage{algpseudocode}
\usepackage{amsmath}
\usepackage{booktabs}
\usepackage[numbers,sort&compress]{natbib}
\usepackage{comment}
\usepackage{arydshln}

\usepackage{etoolbox}
\pretocmd{\maketitle}{%
  \markboth{The definitive version was published in IEEE Transactions on Vehicular Technology, doi: 10.1109/TVT.2026.3710529.}{}%
}{}{}

\newtheorem{remark}{Remark}

\newcommand{\bm}[1]{\mbox{\boldmath{$#1$}}}

\newcommand{\bigcheck}{\scalebox{1.3}{$\boldsymbol{\checkmark}$}}

\begin{document}

\title{Millimeter-Wave Position Sensing Using Reconfigurable Intelligent Surfaces: Positioning Error Bound and Phase Shift Configuration}

\author{Xin Cheng,~\IEEEmembership{Member,~IEEE}, Yuqing Yang, Guangjie Han,~\IEEEmembership{Fellow,~IEEE},
Menglu Li, \\ Ruoguang Li,~\IEEEmembership{Senior Member,~IEEE}, and Feng Shu,~\IEEEmembership{Senior Member,~IEEE}

\thanks{This work was supported in part by the Leading Innovative Talent Introduction and Cultivation Project of Changzhou City under Grant CQ20240137, in part by the Natural Science Foundation of Jiangsu Province under Grant BK20240347, in part by the Hainan Province Science and Technology Special Fund under Grant ZDYF2024GXJS292,  and in part by Hainan Provincial Natural Science Foundation of China under Grant 626ZD0993.  (Corresponding author: Guangjie Han.)}

\thanks{Xin Cheng, Yuqing Yang, Guangjie Han and Ruoguang Li are with the College of Information Science and Engineering, Hohai University, Changzhou 213200, China. (e-mail: xincstar23@163.com).}
\thanks{Menglu Li is with the College of Electrical and Power Engineering, Hohai University, Changzhou 213200, China.}
\thanks{
Feng Shu is with the School of Information and Communication Engineering,
Hainan University, Haikou 570228, China.}}

\maketitle
\begin{abstract}
Millimeter-wave (mmWave) positioning has emerged as a promising technology for next-generation intelligent systems. The advent of reconfigurable intelligent surfaces (RISs) has revolutionized high-precision mmWave localization by enabling dynamic manipulation of wireless propagation environments. This paper investigates a three-dimensional (3D) multi-input single-output (MISO) mmWave positioning system assisted by  multiple RISs. We introduce a measurement framework incorporating sequential RIS activation and directional beamforming to fully exploit virtual line-of-sight (VLoS) paths. The theoretical performance limits are rigorously analyzed through derivation of the subsequent positioning error bound (PEB). To minimize the PEB, the phase shift configurations of the RISs are optimized under both continuous and discrete cases. For continuous phase shifts, the original optimization problem is transformed into a manifold optimization formulation, and a Riemannian manifold-based algorithm is subsequently developed. For discrete phase shifts, a heuristic algorithm is proposed by modifying the grey wolf optimizer. Extensive numerical simulations demonstrate the effectiveness of the proposed algorithms in reducing the PEB and validate the improvement in positioning accuracy achieved by multiple RISs.
\end{abstract}

\IEEEpeerreviewmaketitle

\section{Introduction}
\IEEEPARstart{T}{he} advent of sixth-generation (6G) networks marks a significant leap forward in communication and sensing technology, and promises to revolutionize numerous aspects of modern life  through enhanced connectivity and intelligence \cite{gonzalez2024integrated}. Central to this transformative vision is the capability of precise location awareness. From enabling the seamless operation of smart cities, where infrastructure and services are optimized through real-time location data, to enhancing immersive experiences in augmented reality, where the virtual and physical worlds converge with pinpoint positioning accuracy, the importance of accurate position sensing cannot be overstated.

Millimeter-wave (mmWave) position sensing  stands at the forefront of advancing location-based technology within the realm of 6G networks \cite{chen2022tutorial}. Its significance lies not only in its critical role within the integrated sensing and communication (ISAC) framework of 6G but also in its ability to provide unparalleled precision in positioning by narrow beamwidth and high spatial resolution \cite{chen2022tutorial}.

Traditional millimeter-wave (mmWave) position sensing is achieved by measuring the position-related mmWave characteristics of the line of sight (LoS) path, including received signal strength indicator (RSSI) \cite{vari2014mmwaves}, time of arrival (ToA) and angle of arrival (AoA) \cite{zhou2017low, lin20183, wen20205g}. However, these methods are highly dependent on the presence of LoS path and are particularly susceptible to interference from non-line of sight (NLoS) paths. Due to the high frequency of mmWave signals, the wireless transmission is prone to obstruction by obstacles. To overcome this challenge, utilizing NLoS paths for positioning is also researched \cite{lin2018position,zhou2019successive,wen20205g}.  From a theoretical perspective, the positioning accuracy could be improved. However, the actual estimation is often constrained by the weak signal strength of the uncontrollable NLoS paths compared to the LoS path.  Moreover, accurate  positioning using NLoS paths requires estimating or knowing additional parameters, such as the orientations and positions of the associated scatterers \cite{wen20205g,7102679}.

Recently, reconfigurable intelligent surface (RIS) has been introduced to achieve mmWave position sensing without the LoS path. The RIS consists of an array of low-cost passive reflective elements, each of which can be individually controlled via a feedback loop to modify the electromagnetic wave \cite{cui2014coding,wu2019towards}. In contrast to conventional wireless communication systems that rely on transmit beamforming, RIS-aided systems utilize configurable phase shifts to enable passive beamforming. With properly designed passive beamforming, the RIS is shown to enhance transmission rates, coverage, secure rate and energy efficiency \cite{9769918,9496108,9947328}. Moreover, many of the hardware impairments in practical RIS-assisted wireless communication systems have recently been addressed in the literature \cite{li2024stacked,li2024performance,li2025holographic}. Apart from boosting communication performance, RIS-assisted mmWave systems can also enhance localization precision. With controllable phase shifts, a reflective path can be established if the LoS path is obstructed, which is much stronger than NLoS paths. Moreover, because of extra spatial degrees of freedom (DoFs) introduced by phase shifts, the sensing capability of this system is expected to be high.

\begin{table*}[t]
\centering
\captionsetup{justification=centering, labelsep=newline,textfont=sc}
\caption{Novelty Comparison of Our Paper to the Existing Multiple-RISs-Assisted Positioning Techniques in Literature \cite{ma2024dual,lin2021channel,alhafid2024enhanced,wang2021joint,zhang2023approximate,chen2024multi,liu2022cramer}}
\label{tablecontributions}
\begin{tabular}{c c c c c c c c c}
\hline
 & Our paper &\cite{ma2024dual} & \cite{lin2021channel} & \cite{alhafid2024enhanced} & \cite{wang2021joint} & \cite{zhang2023approximate}  & \cite{chen2024multi} & \cite{liu2022cramer}\\
\hline
PEB-driven phase shift configuration &\bigcheck    & & & & & & \bigcheck & \bigcheck \\
\hdashline
Multi-path propagation environment &\bigcheck   & & \bigcheck& &\bigcheck & \bigcheck &\bigcheck & \\
\hdashline
Exact RIS channel gain &\bigcheck   & & \bigcheck& &\bigcheck &  & & \\
\hdashline
Continuous phase shift &\bigcheck   &\bigcheck &\bigcheck &\bigcheck &\bigcheck &\bigcheck  &\bigcheck &\bigcheck \\
\hdashline
Discrete phase shift &\bigcheck   & & & & &  & & \\
\hline
\end{tabular}
\end{table*}

The sensing ability of RIS has already been studied \cite{9215972,9148744,9124848,cheng2026fingerprint}. When a RIS is involved in a mmWave wireless transmit system, extra position information is provided by  position-related channel parameters of the virtual line-of-sight (VLoS) path.  Normally, the user equipment (UE) position is obtained through exact channel angular and temporal parameters, i.e., AoAs and ToAs based on measurements of received signal. Since the localization performance is changed according to the phase shift of RIS, phase shift design of RIS become the focus of this research \cite{9215972}. A RIS-assisted single-input single-output (SISO) multi-carrier system was investigated in \cite{keykhosravi2021siso,ye2022single}, and the random phase shift design of RIS was studied in \cite{ye2022single}. A RIS-enabled two-dimensional (2D) mmWave multiple output (MIMO) positioning was investigated in \cite{gao2022wireless}, the phase shift design of RIS  and the beamforming of UE are jointly optimized  by minimizing Cramer-Rao lower bound (CRLB)-based error bound, i.e., position error bound (PEB). In \cite{elzanaty2021reconfigurable}, a  RIS is introduced to assist three-dimensional (3D) localization and orientation estimation in a MIMO system. And an analytical phase shift design of RIS was proposed by maximizing the signal-to-noise ratio (SNR) of the reflective path. In \cite{lyu2024crb}, a RIS is explored to serve multiple users and one single target in a  3D downlink mmWave ISAC system. In order to enhance the estimation accuracy as well as the communication performance, the phase shifts of RIS was optimized based on a sensing CRLB minimization problem with a data rate constraint.

To provide more DoFs for position sensing, multiple RISs have been investigated. In \cite{ma2024dual,lin2021channel}, two RISs-assisted position sensing system was studied. Four general phase shift designs including random, structured, grouping and sparse patterns \cite{lin2021channel} and an uniform phase shift design \cite{ma2024dual} were discussed. Apart from the two-RISs structure, general multiple-RISs-assisted sensing systems have also been explored. To eliminate the multiple path interference from multiple reflective paths, authors suggested to differentiate independent reflective paths from each other via one-by-one active method \cite{alhafid2024enhanced,wang2021joint}, an energy-based matching coding method \cite{zhang2023approximate} and time-orthogonal coding method \cite{chen2024multi}. To improve the localization performance, phase shifts of RISs were designed for each reflective path \cite{alhafid2024enhanced,liu2022cramer,wang2021joint,zhang2023approximate,chen2024multi}. An efficient beam sweeping (EBS) scheme was proposed in \cite{alhafid2024enhanced} to scan the area of interest. In the EBS scheme, the phase shifts of RIS were configured to direct a beam towards desired direction at each symbol transmission uniformly. In \cite{liu2022cramer},  the  phase shifts of RIS  are optimized by minimizing PEB  when the free-space scenario is considered.
In \cite{wang2021joint}, a simplified linear RIS structure was considered, and the directional beamforming of AP towards the active RIS and random phase shift design of RIS were used jointly. In \cite{zhang2023approximate}, the phase shifts were optimized by maximizing the gain of reflective path while the UE is assumed to be uniformly distributed in the serve region of RISs. In \cite{chen2024multi}, a multiple-stage phase shift design scheme was developed. Specifically, a random phase shift design was firstly adopted to obtain a rough position estimation of UE, then a directional RIS phase shift design by maximizing the SNR or a derivative RIS phase shift design based on PEB was applied to refine the rough position estimation.

Although the existing literature \cite{ma2024dual,lin2021channel,alhafid2024enhanced,wang2021joint,zhang2023approximate,chen2024multi,liu2022cramer} has conducted relevant research on multiple-RISs-assisted positioning, there still are some shortcomings. First, for the positioning task, the PEB is the most intuitive performance metric, and the phase shifts of RISs should be optimized based on this metric. However, only \cite{chen2024multi,liu2022cramer} have adopted this approach. In other works, phase shifts of RISs are designed either through empirical approaches \cite{lin2021channel,alhafid2024enhanced,wang2021joint} or via SNR-driven approaches \cite{ma2024dual,zhang2023approximate}. Moreover, ideal conditions such as free-space propagation \cite{ma2024dual,alhafid2024enhanced,wang2021joint}, linear RIS structures \cite{liu2022cramer} and simplified RIS channel gain  \cite{ma2024dual,zhang2023approximate,chen2024multi,liu2022cramer} limit general applicabilities. Furthermore, only continuous phase shifts of RISs are considered in these studies.
However, practical implementations often rely on discrete phase shift configurations due to hardware constraints \cite{chen2025integrated}. Investigating discrete phase shifts of RISs is also essential \cite{di2020hybrid}. To bridge the gap in existing research, we initiate a new study focusing on multiple-RISs-assisted mmWave position sensing in this paper. Firstly, the phase shifts of RISs are optimized simultaneously according to the PEB directly. Secondly, the electromagnetic property of planar RIS is considered to express the exact  RIS channel gain. Lastly, both continuous and discrete phase shifts of RISs are investigated. Table \ref{tablecontributions} explicitly contrasts our contributions to the literature at a glance, which are further detailed as follows:
\begin{enumerate}
\item We establish a general 3D mmWave MISO positioning framework assisted by multiple planar RISs and provide theoretical analysis. First, a mmWave signal model is developed, taking into account the electromagnetic property of RIS. Subsequently, a measurement framework with sequential RIS activation scheme and RIS-towards directional beamforming is introduced to sufficiently exploit the VLoS paths enabled by RISs. Moreover, the CRLB and the subsequent PEB  are derived to quantify the system's theoretical positioning limits.
\item To minimize the PEB, two different optimization algorithms are proposed on continuous and discrete phase shifts of RISs, respectively. For the continuous case, a multi-scalar optimization problem is formulated, and it is transformed into a concise single-matrix optimization problem  by theoretical analysis, which provide valuable insight for the phase shift optimization of multiple RISs. Accordingly, an optimization algorithm based on Riemannian manifold is proposed. For the discrete case, a heuristic optimization algorithm based on the grey wolf optimizer is proposed to efficiently search the available solutions, which extends the application of the grey wolf optimizer from continuous to discrete problems.
\item Extensive simulations are conducted for general testing scenarios. The simulation results verify the effectiveness of the proposed algorithms in reducing the PEB for both continuous and discrete configurations. Additionally, key physical factors including the transmitter power, the measurement amount, the number of reflective elements, the RIS position and the RIS amount are analyzed for their impact on the PEB, providing guidance for the system designs.
\end{enumerate}

The remainder of this article is organized as follows: a 3D multiple RISs-assisted mmWave MISO position sensing system is presented in Section II. In Section III, a measurement framework to exploit the VLoS paths is presented, then the corresponding PEB is derived. Subsequently, the optimization algorithms on continuous and discrete phase shifts of RISs are proposed in Section IV. Numerical results are provided in Section V, followed by the conclusions in Section VI.
For convenience, some commonly-used notations are listed in Table \ref{tabecnotations}. The symbols in Table \ref{tabecnotations} are also defined formally where they first appear in the paper.

\emph{Notations:} Boldface lower case and boldface upper case letters denote vectors and matrices, respectively. Sign $(\cdot)^{T}$  denotes the transpose operation while sign $(\cdot)^{-1}$  denotes the inverse operation. Sign $|| \cdot ||$ denotes the Frobenius norm of a matrix or the L2 norm of a vector. Sign $\mathbb{E}\{\cdot\}$ represents the expectation operation. Sign $\mathrm{Tr}(\cdot)$ represents the trace of a matrix. Sign $ (\cdot)^{*} $ denotes the conjugate operation.  Sign $\angle$ represents the angle of a complex number. Sign $\Re$ represents the real part of a matrix. Sign $\odot$ denotes the Hadamard product.
Sign $\mathrm{unit}$ means dividing each element of a matrix by its modulus.

\begin{table}
\centering
\captionsetup{justification=centering, labelsep=newline,textfont=sc}
\caption{Commonly Used Notations}
\label{tabecnotations}
\begin{tabular}{p{0.08\textwidth}|p{0.37\textwidth}}
\hline
\multicolumn{1}{c|}{\textbf{Symbol}} & \multicolumn{1}{c}{\textbf{Definition}} \\
\hline
$\lambda$ & Wavelength \\
\hline
$N_{A}$ & Number of AP antennas \\
\hline
$N$ & Number of measurements \\
\hline
$\bar{L}_{i}$ & Number of reflective elements of the $i$-th RIS  \\
\hline
$d_{x,i}, d_{y,i}$ & Length and width of the reflective element of the $i$-th RIS  \\
\hline
$M_{I,i}, N_{I,i}$ & Row count and column count of the $i$-th RIS  \\
\hline
$\mathbf{a}_{A}$, $\mathbf{a}_{I_{i}}$ & Steering vectors of AP and the $i$-th RIS \\
\hline
$\delta_{0}$, $\delta_{l}$ & Channel gains of the LoS path and the $l$-th NLoS path \\
\hline
$\bar{\delta}_{i}$ & Channel gain of the $i$-th VLoS path with a reflective element \\
\hline
$\vartheta_{i,\bar{l}}$ & Phase shift of the $\bar{l}$-th reflective element \\
\hline
$\mathbf{g}_{i}$, $\mathbf{G}_{i}$ & Phase shift vector and matrix of the $i$-th RIS \\
\hline
$m_{\bar{l},i}$, $n_{\bar{l},i}$ & Column number and row number of the $\bar{l}$-th reflective element of the $i$-th RIS \\
\hline
$G_{A}$, $G_{U}$ & Antenna gains of AP and  UE \\
\hline
$G_{I,i}$ & Scattering gain of the reflective element of the $i$-th RIS \\
\hline
$\mathbf{p}_{A}$, $\mathbf{p}_{U}$, $\mathbf{p}_{I,i}$ & Positions of AP, UE and the $i$-th RIS \\
\hline
$\phi_{AI_{i}}$, $\varphi_{AI_{i}}$ & Elevation angle and azimuth angle from the $i$-th RIS to AP \\
\hline
$\phi_{I_{i}U}$, $\varphi_{I_{i}U}$ & Elevation angle and azimuth angle from the $i$-th RIS to UE \\
\hline
$\mathbf{e}_{x,i}$, $\mathbf{e}_{y,i}$, $\mathbf{e}_{z,i}$ & Horizontal, vertical  and normal direction vectors of the $i$-th RIS \\
\hline
\end{tabular}
\end{table}

\section{System Model}
\begin{figure}[!ht]
  \centering
  \includegraphics[width=0.45\textwidth]{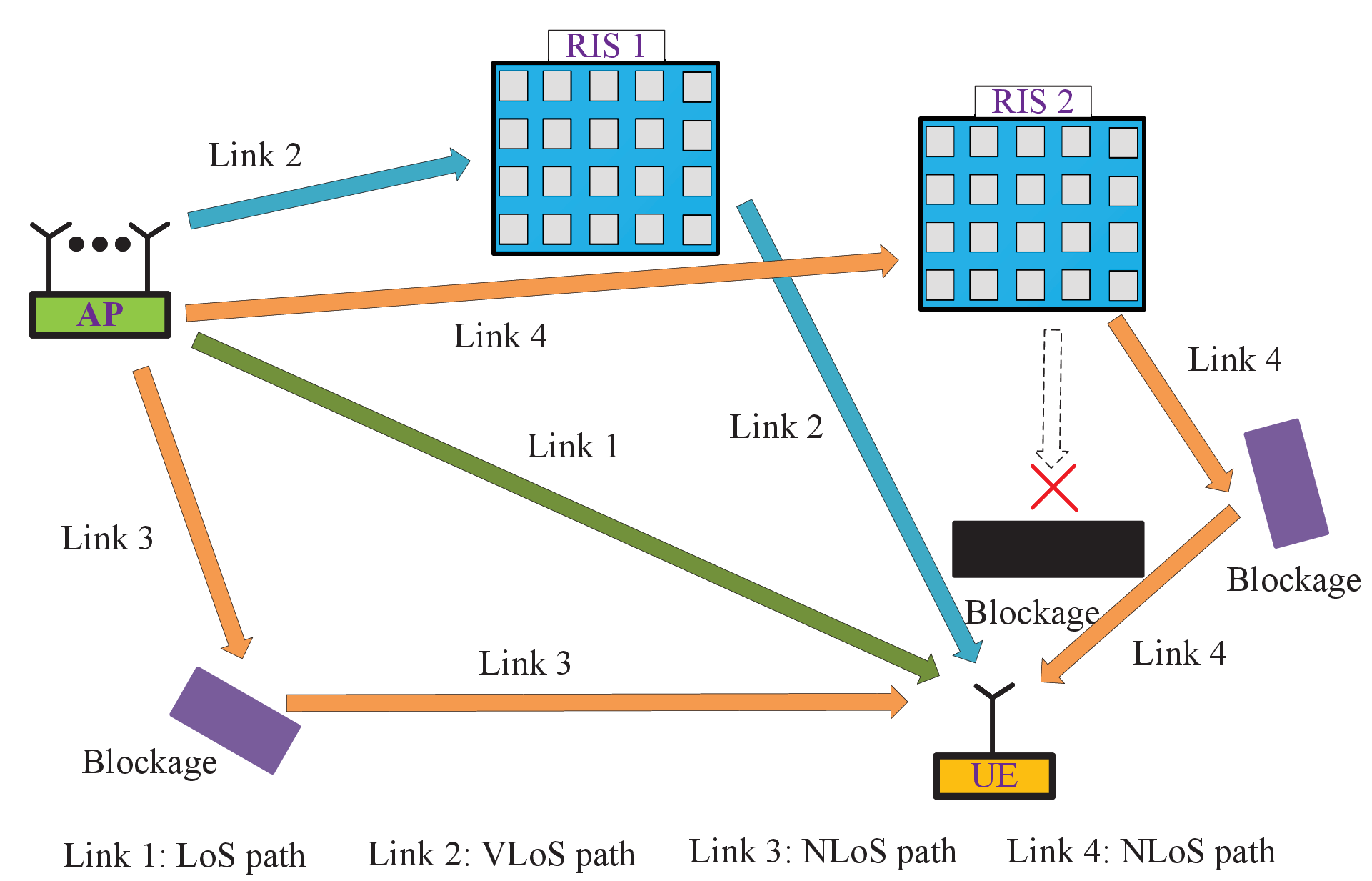}\\
  \caption{RISs-assisted mmWave position sensing system.}\label{sys}
  \vspace{-1\baselineskip}   
\end{figure}

Consider a mmWave position sensing system where the AP adopts an uniform linear array (ULA) antenna structure with $N_{A}$ antennas and the UE equips a single antenna, as shown in Fig.~\ref{sys}.  To guarantee localization servers for the UE as well as reliable linkage between the AP and the UE, $I^{'}$ RISs are deployed, and the AP is able to control RISs via cable or lower frequency radio link. The $i$-th RIS is a rectangular plane, consists of $\bar{L}_{i}$ reflective elements with $N_{I,i}$ rows and $M_{I,i}$ columns. The length of the $i$-th RIS's reflective element is  $d_{x,i}$ while the width is $d_{y,i}$.


The positions of  AP and UE  are denoted as $\mathbf{p}_{A}$ and $\mathbf{p}_{U}$, respectively. The position of the $i$-th RIS is denoted as $\mathbf{p}_{I,i}$. The positions of  AP and RISs are known while the position of UE needs to be estimated by measuring the received mmWave signal transmitted from the AP to the UE with the assistance of RISs. The propagation paths between the AP and the UE consist of three part. One is the LoS path where the signal is directly transmitted from the AP. This path only exists without occlusion. Another is the VLoS path where the signal is transmitted from the AP to the RIS, then transmitted to the UE after reflected by the RIS. The last path refers to the NLoS path where the signal is transmitted from the AP to the UE through reflection, diffraction and scattering caused by  obstacles, e.g., walls, tables, and human bodies\footnote{In this paper, the VLoS path is not involved in the NLoS path. And the path where the signal is reflected by the RIS but also affected by obstacles belongs to the NLoS path.}.

The channel response of the LoS path is expressed as \cite{wang2021joint}
\begin{align}\label{chLoS}
\mathbf{h}_{AU,0}=\delta_{0}\mathbf{a}_{A}^{T}(\theta_{AU}),
\end{align}
where $\delta_{0}$ is the channel  gain of the LoS path,  $\theta_{AU}$ is the angle of departure (AoD) from the AP to the UE,  and $\mathbf{a}_{A}$ represents the steering vector of  AP. According to \cite{tang2022path,tang2020wireless,cheng2021joint}, we have $\delta_{0}=\frac{\sqrt{G_{A}G_{U}}\lambda}{4\pi}d_{AU}^{-1}e^{j2\pi\frac{d_{AU}}{\lambda}}$, where $\lambda$ is the wave strength of transmit signal, $d_{AU}$ is the distance between the AP and the UE with $d_{AU}=\|\mathbf{p}_{A}-\mathbf{p}_{u}\|$, $G_{A}$ is the antenna gain of  AP, and  $G_{U}$ is the antenna gain of UE. Moreover, we have
\begin{align}\label{steerap}
\mathbf{a}_{A}(\theta)=
\begin{bmatrix}e^{j2\pi \alpha_{1}\cos{\theta}} & e^{j2\pi \alpha_{2}\cos{\theta}}
&\cdots & e^{j2\pi \alpha_{N_A}\cos{\theta}}
\end{bmatrix},
\end{align}
where $\alpha_{n}= (\frac{(N_{A}+1)}{2}-n)\frac{\Delta d_{A}}{\lambda},~n=1,2,\cdots,N_{A}$, and $\Delta d_{A}$ is the antenna spacing of AP.

The channel response of the $l$-th NLoS path is expressed as
\begin{align}
\mathbf{h}_{AU,l}=\delta_{l}\mathbf{a}_{A}^{T}(\theta_{AU,l}),
\end{align}
where $\delta_{l}$ is the channel gain of this path with $\delta_{l}\ll \delta_{0}$, $\theta_{AU,l}$ is the AoD of the $l$-th NLoS path, and $L$ is the number of NLoS paths. $\delta_{l}$ and $\theta_{AU,l}$ is determined by the specific wireless propagation environment.

The channel response of the $i$-th VLoS path assisted by the $i$-th RIS is expressed as \cite{wang2021joint}
\begin{align}\label{chVLoS}
\mathbf{h}_{AI_{i}U}=&\bar{\delta}_{i}\mathbf{a}_{I_{i}}^{T}(\phi_{I_{i}U},\varphi_{I_{i}U})\mathbf{G}_{i}
\mathbf{a}_{I_{i}}(\phi_{AI_{i}},\varphi_{AI_{i}})\mathbf{a}_{A}^{T}(\theta_{AI_{i}}),
\end{align}
where $\bar{\delta}_{i}$ is the channel gain of this path with the single reflective element, $\mathbf{a}_{I_{i}}$ is the steering vector of the $i$-th RIS,  $\phi_{I_{i}U}$ and $\varphi_{I_{i}U}$ denote the elevation angle and  azimuth angle at the $i$-th RIS from it to the UE, respectively, $\phi_{AI_{i}}$ and $\varphi_{AI_{i}}$ denote the elevation angle and  azimuth angle at the $i$-th RIS from it to the AP, respectively, and $\mathbf{G}_{i}=\mathrm{diag}\{\mathbf{g}_{i}\}$, denotes the phase shift matrix of the $i$-th RIS.  $\mathbf{g}_{i}$ is the phase shift vector  of the $i$-th RIS. $\mathbf{g}_{i}=\begin{bmatrix} e^{j\vartheta_{i,1}} & e^{j\vartheta_{i,2}} & \cdots&  e^{j\vartheta_{i,\bar{L}_{i}}} \end{bmatrix}$, where $\vartheta_{i,\bar{l}}$, represents the phase shift of the $\bar{l}$-th reflective element with $\bar{l}=1,2,\cdots,\bar{L}_{i}$.

Considering the geometric relationship between the $i$-th RIS and the UE, we have
\begin{subequations}
\begin{align}
\phi_{I_{i}U}=\arccos\left(\frac{(\mathbf{p}_{U}-\mathbf{p}_{I,i})^{T}\mathbf{e}_{z,i}}{ \|\mathbf{p}_{U}-\mathbf{p}_{I,i}\|_{2}}\right),
\end{align}
\begin{align}
\varphi_{I_{i}U}&=\arccos(\cos{\varphi_{I_{i}U}})\\\nonumber
&+(1-\mathrm{sign}(\sin(\varphi_{I_{i}U}))(\pi-\arccos(\cos{\varphi_{I_{i}U}})),
\end{align}
\end{subequations}
with
\begin{subequations}\label{reshipangle}
\begin{align}
\cos{\varphi_{I_{i}U}}=\frac{(\mathbf{p}_{U}-\mathbf{p}_{I,i})^{T}\mathbf{e}_{x,i}}{ \|(\mathbf{p}_{U}-\mathbf{p}_{I,i})-  (\mathbf{p}_{U}-\mathbf{p}_{I,i})^{T}\mathbf{e}_{z,i}\mathbf{e}_{z,i}\|_{2}},
\end{align}
\begin{align}
\sin{\varphi_{I_{i}U}}=\frac{(\mathbf{p}_{U}-\mathbf{p}_{I,i})^{T}\mathbf{e}_{y,i}}{ \|(\mathbf{p}_{U}-\mathbf{p}_{I,i})-  (\mathbf{p}_{U}-\mathbf{p}_{I,i})^{T}\mathbf{e}_{z,i}\mathbf{e}_{z,i}\|_{2}},
\end{align}
\end{subequations}
where $\mathbf{e}_{x,i}$ represents the horizontal direction of the $i$-th RIS, $\mathbf{e}_{y,i}$ represents the vertical direction of the $i$-th RIS, and $\mathbf{e}_{z,i}$ represents the normal direction of the $i$-th RIS.

Considering the physical and electromagnetic property of RIS \cite{tang2020wireless,cheng2021joint,tang2022path}, we have
\begin{align}\label{ampRIS}
\bar{\delta}_{i}=|\bar{\delta}_{i}|e^{j2\pi\frac{d_{AI_{i}}+d_{I_{i}U}}{\lambda}},
\end{align}
with
\begin{align}\label{ampRIS}
&|\bar{\delta}_{i}|\\\nonumber
&=\frac{\sqrt{G_{A}G_{U}G_{I,i}d_{x,i}d_{y,i}F_{i}(\phi_{AI_{i}},\varphi_{AI_{i}})F_{i}(\phi_{I_{i}U},\varphi_{I_{i}U})} A_{i}\lambda}{8\pi^{\frac{3}{2}}d_{AI_{i}}d_{I_{i}U}}\\\nonumber
&=\frac{\sqrt{G_{A}G_{U}F_{i}(\phi_{AI_{i}},\varphi_{AI_{i}})F_{i}(\phi_{I_{i}U},\varphi_{I_{i}U})}d_{x,i}d_{y,i}A_{i}}{4\pi d_{AI_{i}}d_{I_{i}U}},
\end{align}
where $G_{I,i}=\frac{4\pi d_{x,i}d_{y,i}}{\lambda^2}$, represents the scattering gain of a reflective element of the $i$-th RIS \cite{tang2022path}, $A_{i}\leq 1$, represents the uniform amplitude gain of  reflective elements of the $i$-th RIS and $F_{i}$ denotes the normalized power radiation pattern of the $i$-th RIS, and $d_{AI_{i}}$ is the distance between  the AP and the $i$-th RIS with $d_{AI_{i}}=\|\mathbf{p}_{A}-\mathbf{p}_{I,i}\|$ while $d_{I_{i}U}$ is the distance between  the $i$-th RIS and the UE with $d_{I_{i}U}=\|\mathbf{p}_{I,i}-\mathbf{p}_{u}\|$. The directional RIS is considered here. Therefore, for any incidence signal or reflection signal at the $i$-th RIS with elevation angle $\phi$ and azimuth angle $\varphi$, the normalized power radiation pattern is given by
\begin{align}\label{DeF}
F_{i}(\phi,\varphi)=\left\{
\begin{aligned}
&1  ~~~~ \phi\in(0,\frac{\pi}{2}],~~~\varphi\in(0,2\pi] \\
&0 ~~~~ \phi\in(\frac{\pi}{2},\pi],~~~\varphi\in(0,2\pi]
\end{aligned}
\right.,
\end{align}
It  implies that the $i$-th RIS can only control the electromagnetic wave coming from the front side of it in this scenario.

The steering vector of the $i$-th RIS from it to the UE is given by \cite{tang2020wireless}
\begin{align}\label{steerRIS}
&\mathbf{a}_{I_{i}}(\phi_{I_{i}U},\varphi_{I_{i}U})= \\ \nonumber
&~~~~\begin{bmatrix} e^{j2\pi\frac{\Delta d_{I_{i}U,1}}{\lambda}} & e^{j2\pi\frac{\Delta d_{I_{i}U,2}}{\lambda}} & \cdots&  e^{j2\pi\frac{\Delta d_{I_{i}U,\bar{L}_{i}}}{\lambda}}\end{bmatrix},
\end{align}
where $\Delta d_{I_{i}U,\bar{l}}=d_{I_{i}U,\bar{l}}-d_{I_{i}U}$,  $d_{I_{i}U,\bar{l}}$ is the distance between the $\bar{l}$-th reflective element of the $i$-th RIS and the UE. According to the near-field approximation \cite{ling2024channel,pan2023ris},
which is used to compensate the phase variations caused by the distance differences from Tx/Rx to different  reflective elements of RIS, $\Delta d_{I_{i}U,\bar{l}}$  is expressed as
\begin{align}\label{deltadRIS}
\Delta d_{I_{i}U,\bar{l}}\approx&-\Phi_{I_{i}U}\Delta x_{I_{i},\bar{l}}-\Psi_{I_{i}U}\Delta y_{I_{i},\bar{l}}
+\left(1-\Phi^{2}_{I_{i}U}\right)\frac{\Delta x^2_{I_{i},\bar{l}}}{2d_{I_{i}U}} \\\nonumber
&+\left(1-\Psi^{2}_{I_{i}U}\right)\frac{\Delta y^2_{I_{i},\bar{l}}}{2d_{I_{i}U}}-\Phi_{I_{i}U}\Psi_{I_{i}U}\frac{\Delta x_{I_{i},\bar{l}}\Delta y_{I_{i},\bar{l}}}{d_{I_{i}U}},
\end{align}
where $\Delta x_{I_{i},\bar{l}}$ is the horizontal distance between the $\bar{l}$-th reflective element and the center of the $i$-th  RIS, $\Delta y_{I_{i},\bar{l}}$ is the  vertical distance between them,
$\Phi_{I_{i}U}$ and $\Psi_{I_{i}U}$ are defined symbol for simplicity. We have $\Phi_{I_{i}U}=\sin{\phi_{I_{i}U}}\cos{\varphi_{I_{i}U}}$, $\Psi_{I_{i}U}=\sin{\phi_{I_{i}U}}\sin{\varphi_{I_{i}U}}$,
$\Delta x_{I_{i},\bar{l}}=\left(m_{\bar{l},i}-\frac{M_{I,i}+1}{2}\right)d_{x,i}$ and $\Delta y_{I_{i},\bar{l}}=\left(n_{\bar{l},i}-\frac{N_{I,i}+1}{2}\right)d_{y,i}$, where $m_{\bar{l},i}$ denotes the column number of the $\bar{l}$-th reflective element of the $i$-th RIS and $n_{\bar{l},i}$ denotes the row number of it. The derivation of Eq.~(\ref{deltadRIS}) is given in  Appendix A.

The steering vector of the $i$-th RIS from it to the AP is given by \cite{tang2020wireless,cheng2021joint,tang2022path}
\begin{align}
&\mathbf{a}_{I_{i}}(\phi_{AI_{i}},\varphi_{AI_{i}})= \\ \nonumber
&~~~~\begin{bmatrix} e^{j2\pi\frac{\Delta d_{AI_{i},1}}{\lambda}} & e^{j2\pi\frac{\Delta d_{AI_{i},2}}{\lambda}} & \cdots&  e^{j2\pi\frac{\Delta d_{AI_{i},\bar{L}_{i}}}{\lambda}}\end{bmatrix},
\end{align}
where $\Delta d_{AI_{i},\bar{l}}=d_{AI_{i},\bar{l}}-d_{AI_{i}}$,  $d_{AI_{i},\bar{l}}$ is the distance between the $\bar{l}$-th reflective element of the $i$-th RIS and the AP. According to the near-field approximation \cite{ling2024channel,pan2023ris}, $\Delta d_{AI_{i},\bar{l}}$  is expressed as
\begin{align}\label{deltadRIS2}
\Delta d_{AI_{i},\bar{l}}\approx&-\Phi_{AI_{i}}\Delta x_{I_{i},\bar{l}}-\Psi_{AI_{i}}\Delta y_{I_{i},\bar{l}}
+\left(1-\Phi^{2}_{AI_{i}}\right)\frac{\Delta x^2_{I_{i},\bar{l}}}{2d_{AI_{i}}} \\\nonumber
&+\left(1-\Psi^{2}_{AI_{i}}\right)\frac{\Delta y^2_{I_{i},\bar{l}}}{2d_{AI_{i}}}-\Phi_{AI_{i}}\Psi_{AI_{i}}\frac{\Delta x_{I_{i},\bar{l}}\Delta y_{I_{i},\bar{l}}}{d_{AI_{i}}},
\end{align}
where $\Phi_{AI_{i}}$ and $\Psi_{AI_{i}}$ are defined symbol for simplicity. We have $\Phi_{AI_{i}}=\sin{\phi_{AI_{i}}}\cos{\varphi_{AI_{i}}}$, $\Psi_{I_{i}U}=\sin{\phi_{AI_{i}}}\sin{\varphi_{AI_{i}}}$.
The derivation of Eq.~(\ref{deltadRIS2}) is similar to Eq.~(\ref{deltadRIS}), thus omitted here.

$\Delta d_{AI_{i},\bar{l}}$ is the relative distance between the $\bar{l}$-th reflective element of RIS and the AP, expressed as
\begin{align}
\Delta d_{AI_{i},\bar{l}}\approx&-\sin\phi_{AI_{i}}\cos\varphi_{AI_{i}}\left(m_{\bar{l},i}-\frac{M_{I,i}+1}{2}\right)d_{x,i}  \\\nonumber
&-\sin\phi_{AI_{i}}\sin\varphi_{AI_{i}}\left(n_{\bar{l},i}-\frac{N_{I,i}+1}{2}\right)d_{y,i}.
\end{align}

Let $P_{0}$ denote the transmit power of AP, $x$ denote the transmit signal of AP  and $\mathbf{f}$ denote the beamforming of AP.
$\mathbf{f}\in \mathbb{R}^{N_{A}\times1}$, consists of phase-only complex variables, and $\|\mathbf{f}\|=1$. Then the received signal at UE, denoted as $y$, is given by
\begin{align}\label{reeq}
y=&\zeta_{0}\sqrt{P_{0}}\mathbf{h}_{AU,0}\mathbf{f}x \\\nonumber
&+\sum_{i=1}^{I^{'}}\zeta_{i}\gamma_{i}\sqrt{P_{0}}\mathbf{h}_{AI_{i}U}\mathbf{f}x+\sum_{l=1}^{L}\sqrt{P_{0}}\mathbf{h}_{AU,l}\mathbf{f}x+\omega,
\end{align}
where $\zeta_{0}$ denotes the availability indicator of the LoS path, $\zeta_{i}$ is the availability indicator of the VLoS path through the $i$-th RIS, $\gamma_{i}$ denotes the activation status of the $i$-th RIS configured by the AP, $L$ is the number of NLoS paths,  $\mathbf{h}_{AU,0}$ is the channel response of the LoS path, $\mathbf{h}_{AI_{i}U}$ is the channel response of the VLoS path through the $i$-th RIS, $\mathbf{h}_{AU,l}$ is the channel response of the $l$-th NLoS path, and $\omega$ is the zero-mean complex Gaussian additive noise with $w\in \mathcal{CN}(0,\sigma_{\omega})$. It shows that
$\zeta_{0}\in\{0,1\}$, $\zeta_{i}\in\{0,1\}$ and $\gamma_{i}\in\{0,1\}$.

\begin{remark}
It is worth noting that the near-field approximation employs the parabolic wave assumption, which aligns more closely with the actual spherical propagation principle of electromagnetic waves, compared to the planar wave assumption used in the far-field approximation \cite{ling2024channel,cheng2021joint}. Even under far-field conditions where the transmission distance from  RIS is sufficiently large, the near-field approximation remains more accurate than its far-field counterpart. Consequently, the system model in this paper is applicable to both far-field and near-field scenarios.
\end{remark}

\section{Measurement Framework and Error Bound}\label{mfandeb}

\subsection{Measurement Framework}
In order to estimate the position of UE, multiple pilot signals are transmitted form the AP, and the received signals are recorded as the measurement data. As shown in Eq.~(\ref{reeq}), the received signal at UE comes from the LoS path, the VLoS path and the NLoS path. As NLoS path component is uncontrollable and varies fast, it is hard to get useful information from it \cite{wang2021joint}. Differently, the LoS path and the VLoS path can be utilized for position sensing, since the channel responses of the LoS path and the VLoS path can be expressed by the position-related information of UE clearly, as shown in Eq.~(\ref{chLoS}) and  Eq.~(\ref{chVLoS}). However, the LoS path is usually blocked in the complex wireless propagation environment. Fortunately, at least one VLoS path is available through the meticulous deployment of multiple RISs. Therefore, considering the universality, we intend to estimate the position of UE from the VLoS path. It is worth to mention that the received signal from the $i$-th VLoS path contains the position-related information $\phi_{I_{i}U}$, $\varphi_{I_{i}U}$ and $d_{I_{i}U}$, according to Eq.~(\ref{chVLoS}), Eq.~(\ref{ampRIS}) and Eq.~(\ref{steerRIS}).

Assume that $I$ VLoS paths are available for the UE to be estimated, i.e., $\zeta_{i}=1,~i=1,2,\cdots,I$, $\zeta_{i}=0,~I\leq i\leq I^{'}$. The estimation of these availability indicators of VLoS paths has already been proposed in existing works like \cite{wang2021joint}, thus it is assumed as the prior knowledge. To estimate the path-related position information, a sequence measurement diagram \cite{wang2021joint} is applied where all available VLoS paths are measured sequently and independently.

Within the measurement for the $i$-th available VLoS path, the $i$-th RIS is active while the other RISs are inactive, i.e., $\gamma_{i}=1$,  $\gamma_{j}=0,~j=1,2\cdots,I,~j\neq i$, and $N$ pilot signals with $x=1$ are transmitted from the AP at different time slot. Then, the received signal at the $n$-th time slot is expressed as
\begin{align}
&y_{i,n}\\\nonumber
&=\zeta_{0}\sqrt{P_{0}}\mathbf{h}_{AU,0}\mathbf{f}_{i,n}+\sqrt{P_{0}}\mathbf{h}_{AI_{i}U}\mathbf{f}_{i,n}\\\nonumber
&+\sum_{l=1}^{L}\sqrt{P_{0}}\mathbf{h}_{AU,l}\mathbf{f}_{i,n}+\omega_{i,n}\\\nonumber
&=\sqrt{P_{0}}\bar{\delta}_{i}\mathbf{a}_{I_{i}}^{T}(\phi_{I_{i}U},\varphi_{I_{i}U}) \mathbf{G}_{i,n}\mathbf{a}_{I_{i}}(\phi_{AI_{i}},\varphi_{AI_{i}})\mathbf{a}_{A}^{T}(\theta_{AI_{i}})\mathbf{f}_{i,n}\\\nonumber
&+\zeta_{0}\underbrace{\sqrt{P_{0}}\delta_{0}\mathbf{a}_{A}^{T}(\theta_{AU})\mathbf{f}_{i,n}}_{\varrho_{i,n}}
+\underbrace{\sqrt{P_{0}}\sum_{l=1}^{L}\delta_{l}\mathbf{a}_{A}^{T}(\theta_{AU,l})\mathbf{f}_{i,n}}_{v_{i,n}}
+\omega_{i,n} \\\nonumber
&=\sqrt{P_{0}}\bar{\delta}_{q}(\mathbf{a}_{I_{i}}(\phi_{I_{i}U},\varphi_{I_{i}U})\odot\mathbf{a}_{I_{i}}(\phi_{AI_{i}},\varphi_{AI_{i}}))^{T}\\\nonumber
&~~~~~~\mathbf{g}_{i,n}\mathbf{a}_{A}^{T}(\theta_{AI_{i}})\mathbf{f}_{i,n}+\zeta_{0}\varrho_{i,n}+v_{i,n}+\omega_{i,n},
\end{align}
where $\mathbf{G}_{i,n}$ and $\mathbf{g}_{i,n}$ denotes the phase shift matrix and phase shift vector of the $i$-th RIS at the $n$-th time slot, respectively. It implies that besides the received noise, there are two interference terms, caused by the LoS path and the NLoS path. Generally, due to the small scale and randomness of $\delta_{l}$, $v_{i,n}$ follows a complex Gaussian distribution, i.e.,  $v_{i,n}\in \mathcal{CN}(0,\sigma_{v,i}^2)$.  Moreover, $w_{i,n}$ is the additive receive noise, and $\omega_{i,n}\in \mathcal{CN}(0,\sigma_{\omega}^2)$.

To get the useful information from the VLoS path, it is useful to concentrate the power of AP towards the $i$-th RIS via beamforming \cite{wang2021joint}. Mathematically, the beamforming is designed as
\begin{align}\label{beamRISq}
\mathbf{f}_{i,n}=\frac{\mathbf{a}^{*}_{A}(\theta_{AI_{i}})}{\sqrt{N_{A}}}.
\end{align}
In this way, the interference term from the LoS path becomes insignificant, i.e., $\varrho_{i,n}\approx0$. This is caused by the spatial filtering impact, i.e., $\mathbf{a}_{A}^{T}(\theta_{AU})\mathbf{a}_{A}^{*}(\theta_{AI_{i}})\approx0$ for $|\theta_{AU}-\theta_{AI_{i}}|>\frac{1}{N_{A}}$.

From above, the received signal at the $n$-th time slot is written as
\begin{align}\label{cmvlosn}
&y_{i,n}=\\\nonumber
&\underbrace{\sqrt{N_{A}P_{0}}\bar{\delta}_{i}(\mathbf{a}_{I_{i}}(\phi_{I_{i}U},\varphi_{I_{i}U})\odot\mathbf{a}_{I_{i}}(\phi_{AI_{i}},\varphi_{AI_{i}}))^{T}\mathbf{g}_{i,n}}_{\mu_{i,n}}
+\eta_{i,n},
\end{align}
where $\eta_{i,n}=v_{i,n}+w_{i,n}$, represents the integrated interference and noise, and $\eta_{i,n}\in \mathcal{CN}(0,\sigma^2_{\eta_{i}})$ where $\sigma^2_{\eta_{i}}=\sigma_{w}^2+\sigma_{v,i}^2$. After $N$ time slots, i.e., $N$ times of measurement, the received signals are collected in the following vector form.
\begin{align}\label{cmvlos}
&\mathbf{y}_{i}=\\\nonumber
&\sqrt{N_{A}P_{0}}\bar{\delta}_{i}(\mathbf{a}_{I_{i}}(\phi_{I_{i}U},\varphi_{I_{i}U})\odot\mathbf{a}_{I_{i}}(\phi_{AI_{i}},\varphi_{AI_{i}}))^{T}\mathbf{\bar{G}}_{i}
+\bm{\eta}_{i}, \\\nonumber
\end{align}
where $\mathbf{y}_{i}=\begin{bmatrix} y_{i,1} & y_{i,2} & \cdots&  y_{i,N} \end{bmatrix}^{T}$, $\mathbf{\bar{G}}_{i}= \begin{bmatrix} \mathbf{g}_{i,1}^{T} & \mathbf{g}_{i,2}^{T} & \cdots&  \mathbf{g}_{i,N}^{T} \end{bmatrix}^{T}$ and $\bm{\eta}_{i}=\begin{bmatrix} \eta_{i,1} & \eta_{i,2} & \cdots&  \eta_{i,N} \end{bmatrix}^{T}$ where $\bm{\eta}_{i}\in \mathcal{CN}(0,\sigma^2_{\eta_{i}}I)$.

After measuring the $I$ VLoS paths sequently, the complete measurement data is collected, expressed as $\{\mathbf{y}_{i}\}_{i=1}^{I}$. Based on $\{\mathbf{y}_{i}\}_{i=1}^{I}$, the position of UE can be estimated via maximum likelihood estimate (MLE). Many numerical algorithms can be used to solve  such MLE problem such as grid search and Taylor-series method.

\subsection{Error Bound}
It is worth to mention that $\mathbf{y}_{i}$ contains the position-related information $\phi_{I_{i}U}$, $\varphi_{I_{i}U}$ and $d_{I_{i}U}$, which reflects the position of UE in the 3D indoor space.  Let $\bm{\tau}_{i}\triangleq\begin{bmatrix} \phi_{I_{i}U} & \varphi_{I_{i}U} \end{bmatrix}$,  $\bm{\xi}_{i}\triangleq\begin{bmatrix} \phi_{I_{i}U} & \varphi_{I_{i}U} &  d_{I_{i}U} \end{bmatrix}$, $\bm{\tau}\triangleq\begin{bmatrix} \bm{\tau_{1}} & \bm{\tau_{2}} &\cdots & \bm{\tau_{I}} \end{bmatrix}$ and $\bm{\xi}\triangleq\begin{bmatrix} \bm{\xi_{1}} & \bm{\xi_{2}} &\cdots & \bm{\xi_{I}} \end{bmatrix}$.

Let  $\mathbf{F}_{\pmb{\xi}_{i}}$ denote the Fisher information matrix (FIM) about the position-related information from $\mathbf{y}_{i}$. Based on Eq.~(\ref{cmvlos}), and referring to \cite{kay1993fundamentals}, we have
\begin{align}\label{FIMVLoSii}
\mathbf{F}_{\pmb{\xi}_{i}}(a,b)=\frac{2}{\sigma_{w}^2+\sigma_{v}^2} \sum_{n=1}^{N}\Re\left\{\frac{\partial \mu_{i,n}^{*}}{\partial \bm{\xi}_{i}(a)} \frac{\partial \mu_{i,n}}{ \partial \bm{\xi}_{i}(b)}\right\},~a,b=1,2,3.
\end{align}
The related derivations are given as follows.
\begin{align}
&\frac{\partial \mu_{i,n}}{\partial \bm{\xi}_{i}(1)}\\\nonumber
&=\sqrt{N_{A}P_{0}}\bar{\delta}_{i} \frac{\partial (\mathbf{a}_{I_{i}}(\phi_{I_{i}U},\varphi_{I_{i}U})\odot\mathbf{a}_{I_{i}}(\phi_{AI_{i}},\varphi_{AI_{i}}))^{T}}{\partial \phi_{I_{i}U}}\mathbf{g}_{i,n}\\\nonumber
&=\sqrt{N_{A}P_{0}}\bar{\delta}_{i} (\mathbf{b}_{I_{i}U,\phi}\odot \mathbf{a}_{AI_{i}U})^{T}\mathbf{g}_{i,n},
\end{align}
where $\mathbf{a}_{AI_{i}U}\triangleq\mathbf{a}_{I_{i}}(\phi_{I_{i}U},\varphi_{I_{i}U})\odot\mathbf{a}_{I_{i}}(\phi_{AI_{i}},\varphi_{AI_{i}})$, and  $\mathbf{b}_{I_{i}U,\phi}\in \mathbb{R}^{\bar{L}_{i}\times1}$, and the $\bar{l}$-th element of it is written as
\begin{align}
&\mathbf{b}_{I_{i}U,\phi}(\bar{l})=j2\pi\frac{1}{\lambda}\frac{\partial \Delta d_{I_{i}U,\bar{l}}}{\partial \phi_{I_{i}U}} \\\nonumber
&=-j2\pi\cos{\phi_{I_{i}U}}\cos{\varphi_{I_{i}U}}\frac{\Delta x_{I_{i},\bar{l}}}{\lambda}  \\\nonumber
&~~~-j2\pi\cos{\phi_{I_{i}U}}\sin{\varphi_{I_{i}U}}\frac{\Delta y_{I_{i},\bar{l}}}{\lambda}\\\nonumber
&~~~-j2\pi\sin{2\phi_{I_{i}U}}\cos^2{\varphi_{I_{i}U}}\frac{\Delta x^2_{I_{i},\bar{l}}}{2d_{I_{i}U}\lambda}\\\nonumber
&~~~-j2\pi\sin{2\phi_{I_{i}U}}\sin^2{\varphi_{I_{i}U}}\frac{\Delta y^2_{I_{i},\bar{l}}}{2d_{I_{i}U}\lambda}\\\nonumber
&~~~-j2\pi\sin{2\phi_{I_{i}U}}\cos{\varphi_{I_{i}U}}\sin{\varphi_{I_{i}U}}\frac{\Delta x_{I_{i},\bar{l}}\Delta y_{I_{i},\bar{l}}}{d_{I_{i}U}\lambda}.
\end{align}

Similarly, we have
\begin{align}
\frac{\partial \mu_{i,n}}{\partial \bm{\xi}_{i}(2)}=\sqrt{N_{A}P_{0}}\bar{\delta}_{i} (\mathbf{b}_{I_{i}U,\varphi}\odot \mathbf{a}_{AI_{i}U})^{T}\mathbf{g}_{i,n},
\end{align}
where $\mathbf{b}_{I_{i}U,\varphi}\in \mathbb{R}^{\bar{L}_{i}\times1}$, and the $\bar{l}$-th element of it is written as
\begin{align}
&\mathbf{b}_{I_{i}U,\varphi}(\bar{l})
=j2\pi\frac{1}{\lambda}\frac{\partial \Delta d_{I_{i}U,\bar{l}}}{\partial \varphi_{I_{i}U}} \\\nonumber
&=j2\pi\sin{\phi_{I_{i}U}}\sin{\varphi_{I_{i}U}}\frac{\Delta x_{I_{i},\bar{l}}}{\lambda}  \\\nonumber
&~~-j2\pi\sin{\phi_{I_{i}U}}\cos{\varphi_{I_{i}U}}\frac{\Delta y_{I_{i},\bar{l}}}{\lambda}\\\nonumber
&~~~+j2\pi\sin^2{\phi_{I_{i}U}}\sin{2\varphi_{I_{i}U}}\frac{\Delta x^2_{I_{i},\bar{l}}}{2d_{I_{i}U}\lambda}\\\nonumber
&~~~-j2\pi\sin^2{\phi_{I_{i}U}}\sin{2\varphi_{I_{i}U}}\frac{\Delta y^2_{I_{i},\bar{l}}}{2d_{I_{i}U}\lambda}\\\nonumber
&~~~-j2\pi\sin^2{\phi_{I_{i}U}}\cos{2\varphi_{I_{i}U}}\frac{\Delta x_{I_{i},\bar{l}}\Delta y_{I_{i},\bar{l}}}{d_{I_{i}U}\lambda}.
\end{align}

Lastly, we have
\begin{align}
\frac{\partial\mu_{i,n}}{\partial \bm{\xi}_{i}(3)}&=\sqrt{N_{A}P_{0}}\frac{\partial \bar{\delta}_{i}\mathbf{a}_{AI_{i}U}^{T}}{\partial d_{I_{i}U}}\mathbf{g}_{i,n}\\\nonumber
&=\sqrt{N_{A}P_{0}}\left(\frac{\partial \bar{\delta}_{i}}{\partial d_{I_{i}U}}\mathbf{a}_{AI_{i}U}^{T}+\bar{\delta}_{i}\frac{\partial \mathbf{a}_{AI_{i}U}^{T}}{\partial d_{I_{i}U}}  \right)\mathbf{g}_{i,n}\\\nonumber
&=\sqrt{N_{A}P_{0}}\bar{\delta}_{i}(\mathbf{b}_{I_{i}U,d}\odot\mathbf{a}_{AI_{i}U})^{T}\mathbf{g}_{i,n},
\end{align}
where $\mathbf{b}_{I_{i}U,d}\in \mathbb{R}^{\bar{L}_{i}\times1}$, and the $\bar{l}$-th element of it is written as
\begin{align}
&\mathbf{b}_{I_{i}U,d}(\bar{l})=-\frac{1}{d_{I_{i}U}} +j2\pi\frac{1}{\lambda}-j2\pi\frac{1}{\lambda}\left(1-\Phi^{2}_{I_{i}U}\right)\frac{\Delta x^2_{I_{i},\bar{l}}}{2d^2_{I_{i}U}}\\\nonumber
&-j2\pi\frac{1}{\lambda}\left(1-\Psi^{2}_{I_{i}U}\right)\frac{\Delta y^2_{I_{i},\bar{l}}}{2d^2_{I_{i}U}}+j2\pi\frac{1}{\lambda}\Phi_{I_{i}U}\Psi_{I_{i}U}\frac{\Delta x_{I_{i},\bar{l}}\Delta y_{I_{i},\bar{l}}}{d^2_{I_{i}U}}.
\end{align}

Let  $\mathbf{F}_{\pmb{\xi}}$ denote the FIM about the position-related channel parameters from $\{\mathbf{y}_{i}\}_{i=1}^{I}$.  Referring to \cite{kay1993fundamentals}, we have
\begin{align}
\mathbf{F}_{\pmb{\xi}}=\mathrm{diag}(\mathbf{F}_{\pmb{\xi}_{1}},\mathbf{F}_{\pmb{\xi}_{2}},\cdots,\mathbf{F}_{\pmb{\xi}_{I}}).
\end{align}

Note that there are nonlinear relationships between $\pmb{\xi}$ and the position of UE, as shown in Eq.~(\ref{reshipangle}). To derive the FIM about the position of UE, we define the following transformation matrix.
\begin{align}
\mathbf{R}\triangleq\frac{\partial  \bm{\xi}}{\partial \mathbf{p}_{U}}=\begin{bmatrix} \frac{\partial \pmb{\xi}_{1}}{\partial \mathbf{p}_{U}}^{T} &  \frac{\partial \pmb{\xi}_{2}}{\partial \mathbf{p}_{U}}^{T} & \cdots & \frac{\partial \pmb{\xi}_{I}}{\partial \mathbf{p}_{U}}^{T}\end{bmatrix}^{T}.
\end{align}
And the elements of the transformation matrix are given as follows.
\begin{align}\label{trsele1}
&\frac{\partial \bm{\xi}_{i}(1)}{\partial \mathbf{p}_{U}}=\frac{\partial\arccos{\cos{\phi_{I_{i}U}}}}{\partial \mathbf{p}_{U}}
=\frac{-1}{\sqrt{1-\cos^2{\phi_{I_{i}U}}}}\frac{\partial \cos{\phi_{I_{i}U}}}{\partial \mathbf{p}_{U}}\\\nonumber
&=\frac{\|\mathbf{p}_{U}-\mathbf{p}_{I,i}\|\mathbf{e}_{z,i}-(\mathbf{p}_{U}-\mathbf{p}_{I,i})^{T}\mathbf{e}_{z,i}\frac{\mathbf{p}_{U}-\mathbf{p}_{I,i}}{\|\mathbf{p}_{U}-\mathbf{p}_{I,i}\|}    }{(\|\mathbf{p}_{U}-\mathbf{p}_{I,i}\|^{2}-((\mathbf{p}_{U}-\mathbf{p}_{I,i})^{T}\mathbf{e}_{z,i})^2))^{\frac{1}{2}}\|\mathbf{p}_{U}-\mathbf{p}_{I,i}\|}.
\end{align}
\begin{align}\label{trsele2}
&\frac{\partial \bm{\xi}_{i}(2)}{\partial \mathbf{p}_{U}}=\mathrm{sign}(\sin{\varphi_{I_{i}U}})\frac{\partial\arccos{\cos{\varphi_{I_{i}U}}}}{\partial \mathbf{p}_{U}}\\\nonumber
&=\frac{-\mathrm{sign}(\sin{\varphi_{I_{i}U}})}{\sqrt{1-\cos^2{\varphi_{I_{i}U}}(\mathbf{p}_{U})}}\frac{\partial \cos{\varphi_{I_{i}U}}(\mathbf{p}_{U})}{\partial \mathbf{p}_{U}}\\\nonumber
&=\mathrm{sign}((\mathbf{p}_{U}-\mathbf{p}_{I,i})^{T}\mathbf{e}_{y,i})\\\nonumber
&~~\frac{(\mathbf{p}_{U}-\mathbf{p}_{I,i})^{T}\mathbf{e}_{x,i} ((\mathbf{p}_{U}-\mathbf{p}_{I,i})-(\mathbf{p}_{U}-\mathbf{p}_{I,i})^{T}\mathbf{e}_{z,i}\mathbf{e}_{z,i})}{|(\mathbf{p}_{U}-\mathbf{p}_{I,i})^{T}\mathbf{e}_{y,i}|   \|(\mathbf{p}_{U}-\mathbf{p}_{I,i})-(\mathbf{p}_{U}-\mathbf{p}_{I,i})^{T}\mathbf{e}_{z,i}\mathbf{e}_{z,i}   \|^{2}}.
\end{align}
\begin{align}\label{trsele3}
\frac{\partial \bm{\xi}_{i}(3)}{\partial \mathbf{p}_{U}}=
\frac{\partial d_{I_{i}U}}{\partial \mathbf{p}_{U}}=\frac{\mathbf{p}_{U}-\mathbf{p}_{I,i}}{ \|\mathbf{p}_{U}-\mathbf{p}_{I,i}\|}.
\end{align}
Eq.~(\ref{trsele1}) is resulted from $\phi_{I_{i}U}= \arccos{\cos{\phi_{I_{i}U}}}$ for $\phi_{I_{i}U}\in (0,\frac{\pi}{2}]$, and Eq.~(\ref{trsele2}) is resulted from $\varphi_{I_{i}U}= \arccos{\cos{\varphi_{I_{i}U}}}+(1-\mathrm{sign}(\sin{\varphi_{I_{i}U}}))(\pi-\arccos{\cos{\varphi_{I_{i}U}}})$
for $\varphi_{I_{i}U}\in (0,2\pi]$.

Let  $\mathbf{F}_{\mathbf{p}_{U}}$ denote the FIM about the position of UE from $\{\mathbf{y}_{i}\}_{i=1}^{I}$. The FIM can be obtained by the chain rule as \cite{elzanaty2021reconfigurable}
\begin{align}\label{martrfinal}
\mathbf{F}_{\mathbf{p}_{U}}=\mathbf{R}\mathbf{F}_{\pmb{\xi}}\mathbf{R}^{T}.
\end{align}
And the root mean square error (RMSE) for any unbiased estimation, denoted as $E_{\mathbf{p}_{U}}$, observes the following inequality.
\begin{align}\label{lbrmse}
E_{\mathbf{p}_{U}}\geq \mathrm{L_R}\triangleq\sqrt{\mathrm{Tr}\{\mathbf{F}_{\mathbf{p}_{U}}^{-1}\}}.
\end{align}
The term $\mathrm{L_R}$ represents the lower bound of RMSE, i.e., PEB.

\section{Phase Shift Optimization}
The PEB implies the effectiveness of measurement, thus is widely used to guide the measurement of localization tasks \cite{ucinski2004optimal}. As implied by $\mathrm{L_R}$, the phase shifts of RISs play an important role in the limited positioning accuracy. To improve it, the phase profile should be properly designed to exploit the potential of RISs as much as possible. In existing studies, the phase shifts of an RIS can be divided into continuous and discrete \cite{chen2025integrated,di2020hybrid}. The continuous phase shift serves as a theoretical benchmark, guiding performance limits and algorithm designs, while the discrete phase shift drives practical innovations, addresses hardware constraints and deployment feasibilities. Therefore, research on both configurations is vital. In the following subsections, optimizations on continuous and discrete phase shifts are analysed, respectively.

\subsection{Continuous Phase Shift Optimization}
The optimization problem on continuous phase shifts can be formed  as follows.
\begin{align}\label{pl}
\mathcal{P}_{1}: &\min_{\{\{\mathbf{g}_{i,n}\}_{n=1}^{N}\}_{i=1}^{I}}~~~~~~~~~~~~~~~~~\mathrm{L_R}    \nonumber\\
&~~~~~~~\text{s. t.}~~~~~~~~~~~~~~~~~|\mathbf{g}_{i,n}(\bar{l})|=1,\\\nonumber
&~~n=1,2,\cdots,N, ~\bar{l}=1,2,\cdots,\bar{L}_{i}, ~i=1,2,\cdots,I.
\end{align}
After obtaining optimized ${\{\{\mathbf{g}_{i,n}\}_{n=1}^{N}\}_{i=1}^{I}}$, the $\bar{l}$-th phase shift of the $i$-th RIS at the $n$-th measurement, denoted as $\vartheta^{n}_{i,\bar{l}}$, is given by
\begin{align}\label{recovephasefinal}
\vartheta^{n}_{i,\bar{l}}=\angle\left\{\mathbf{g}_{i,n}(\bar{l})\right\}.
\end{align}


Problem $\mathcal{P}_{1}$ is cumbersome with multiple vector variables, before solving it,  we transform the form of it so that only one matrix variable is involved.

Firstly, based on Eq.~(\ref{FIMVLoSii}), the FIM related to the $i$-th RIS is expressed as
\begin{align}\label{FIMiiVLoSfo}
\mathbf{F}_{\pmb{\xi}_{i}}&=\frac{2}{\sigma_{w}^2+\sigma_{v}^2} \Re\left\{\sum_{n=1}^{N}\frac{\partial \mu_{i,n}}{\partial \pmb{\xi_{i}}} \left(\frac{\partial \mu_{i,n}}{ \partial \pmb{\xi_{i}}}\right)^{H}\right\}\\\nonumber
&=\frac{2N_{A}P_{0}|\bar{\delta}_{i}|^2}{\sigma_{w}^2+\sigma_{v}^2}\Re\left\{\sum_{n=1}^{N} \pmb{\kappa}\mathbf{g}_{i,n}\mathbf{g}^{H}_{i,n}\pmb{\kappa}^{H}\right\}\\\nonumber
&=\frac{2N_{A}P_{0}|\bar{\delta}_{i}|^2}{\sigma_{w}^2+\sigma_{v}^2}\Re\left\{\pmb{\kappa}\left(\sum_{n=1}^{N}\mathbf{g}_{i,n}\mathbf{g}^{H}_{i,n}\right)\pmb{\kappa}^{H}\right\}\\\nonumber
&=\frac{2N_{A}P_{0}|\bar{\delta}_{i}|^2}{\sigma_{w}^2+\sigma_{v}^2}\Re\left\{\pmb{\kappa}\mathbf{\bar{G}}_{i}\mathbf{\bar{G}}_{i}^{H}\pmb{\kappa}^{H}\right\},
\end{align}
where
\begin{align}
\pmb{\kappa}_{i}=\begin{bmatrix}(\mathbf{b}_{I_{i}U,\phi}\odot \mathbf{a}_{AI_{i}U})^{T}\\
(\mathbf{b}_{I_{i}U,\varphi}\odot \mathbf{a}_{AI_{i}U})^{T}\\
(\mathbf{b}_{I_{i}U,d}\odot\mathbf{a}_{AI_{i}U})^{T}\end{bmatrix},
\end{align}
\begin{align}\label{defineGi}
\mathbf{\bar{G}}_{i}= \begin{bmatrix} \mathbf{g}_{i,1}^{T} & \mathbf{g}_{i,2}^{T} & \cdots&  \mathbf{g}_{i,N}^{T} \end{bmatrix}^{T}.
\end{align}
Note that $\mathbf{\bar{G}}_{i}$ is a  matrix collecting the phase shift vectors of the $i$-th RIS at different measuring time slots.

Using the FIM related to the i-th RIS derived in Eq. (\ref{FIMiiVLoSfo}), we reformulate the collective FIM of all RISs in Eq. (\ref{martrfinal}) as
\begin{align}\label{FIMVLoSfo}
\mathbf{F}_{\mathbf{p}_{U}}&=\sum_{i=1}^{I}\mathbf{R}_{i}\mathbf{F}_{\pmb{\xi}_{i}}\mathbf{R}_{i}^{T}\\\nonumber
&=\frac{2N_{A}P_{0}}{\sigma_{w}^2+\sigma_{v}^2}\sum_{i=1}^{I}|\bar{\delta}_{i}|^2\Re\{\mathbf{R}_{i}\pmb{\kappa}_{i}\mathbf{\bar{G}}_{i}\mathbf{\bar{G}}_{i}^{H}\pmb{\kappa}^{H}_{i}\mathbf{R}^{H}_{i}\}.
\end{align}
Subsequently, problem $\mathcal{P}_{1}$ is equivalently transformed into the following formulation.
\begin{align}\label{pl2}
\mathcal{P}_{2}: &\min_{\mathbf{\{\bar{G}}_{i}\}_{i=1}^{N}}~~\mathrm{Tr}\{(\sum_{i=1}^{I}\Re\{|\bar{\delta}_{i}|^2\mathbf{R}_{i}\pmb{\kappa}_{i}\mathbf{\bar{G}}_{i}\mathbf{\bar{G}}_{i}^{H}\pmb{\kappa}^{H}_{i}\mathbf{R}^{H}_{i}\})^{-1}\}
\nonumber\\
&~~\text{s. t.}~~~~~~~~~~~~~~~~~~|\mathbf{\bar{G}}_{i}(l,n)|=1,\\\nonumber
&~~~~~~~~~~~~~n=1,2,\cdots,N, ~\bar{l}=1,2,\cdots,\bar{L}_{i}.
\end{align}
The optimization problem now involves determining the independent phase shift matrices of different RISs.

To simplify problem $\mathcal{P}_{2}$,  we define a composite phase shift matrix across all RISs as follows.
\begin{align}\label{defineG}
\mathbf{\bar{G}}=\begin{bmatrix} \mathbf{\bar{G}}_{1} &  \mathbf{\bar{G}}_{2} & \cdots & \mathbf{\bar{G}}_{I} \end{bmatrix}.
\end{align}

To recover $\mathbf{\bar{G}}_{i}$ from $\mathbf{\bar{G}}$,  transformation matrices are defined by
\begin{align}
\mathbf{H}_{i}=\begin{bmatrix} (\mathbf{H}^{1}_{i})^{T} & (\mathbf{H}^{2}_{i})^{T}  & \cdots &  (\mathbf{H}^{I}_{i})^{T}  \end{bmatrix}^{T},
\end{align}
with
\begin{align}
\mathbf{H}^{i^{'}}_{i}=\left\{
\begin{aligned}
&\mathbf{I}_{\bar{L}_{i}\times \bar{L}_{i}}  ~~~~ i^{'}=i \\
&\mathbf{0}_{\bar{L}_{i}\times \bar{L}_{i}} ~~~~  i^{'}\neq i
\end{aligned}
\right..
\end{align}
The identity $\mathbf{H}_{i}\mathbf{\bar{G}}=\mathbf{\bar{G}}_{i}$ can be directly verified through matrix multiplication. Then, problem $\mathcal{P}_{2}$ is equivalently transformed into the following formulation.
\begin{align}\label{pl3}
\mathcal{P}_{3}: &\min_{\mathbf{\bar{G}}}~~\mathrm{Tr}\{(\sum_{i=1}^{I}\Re\{|\bar{\delta}_{i}|^2\mathbf{R}_{i}\pmb{\kappa}_{i}\mathbf{H}_{i}\mathbf{\bar{G}}\mathbf{\bar{G}}^{H}\mathbf{H}^{H}_{i}\pmb{\kappa}^{H}_{i}\mathbf{R}^{H}_{i}\})^{-1}\}
\nonumber\\
&~~\text{s. t.}~~~~~~~~~~~~~~~~~~|\mathbf{\bar{G}}(l,n)|=1,\\\nonumber
&~~~~~~~~~n=1,2,\cdots,N, ~\bar{l}=1,2,\cdots,\sum_{i=1}^{I}\bar{L}_{i}.
\end{align}
Problem $\mathcal{P}_{3}$ is also equivalent to problem $\mathcal{P}_{1}$, but more concise to solve.

Problem $\mathcal{P}_{3}$ is challenging to address with the non-linear objective function, the high-dimensional matrix variable and unit-modulus constrains. To address it, manifold optimization is used by treating the high-dimensional matrix variable with unit modulus constraints as an element on the  complex circle manifold. According to the manifold optimization theory \cite{absil2009optimization}, the  complex circle manifold is described by
\begin{align}
\mathcal{O}= \left\{\mathbf{\bar{G}} \in \mathbb{C}^{(\sum_{i=1}^{I}\bar{L}_{i})\times N} \big{|} |\mathbf{\bar{G}}(l,n)|=1, \forall l,n\right\}.
\end{align}
When optimizing $\mathbf{\bar{G}}$ over the complex circle manifold, the unit modulus constraint is inherently satisfied \cite{xu2019resource}. Here, we propose a manifold gradient-based algorithm outlined in Algorithm \ref{alg_1}, which is an iterative algorithm, and similar to the gradient descent algorithm in the Euclidean space.

\begin{algorithm}
\caption{Continuous Phase Shift Optimization Algorithm Based on Riemannian Manifold (CPSOA-RM)}
\label{alg_1}
\begin{algorithmic}[1]
\Require Tolerance $\epsilon$; Scalars for Armijo step size  $m_{\mathrm{max}}$, $\alpha_{A}$, $\beta_{A}$, $\sigma$.
\Ensure  Phase shifts of RISs $\{\{\{\vartheta^{n}_{i,\bar{l}}\}_{n=1}^{N}\}_{\bar{l}=1}^{\bar{L}_{i}}\}_{i=1}^{I}$.
\State \textbf{Initialize:} $\mathbf{\bar{G}}^{0}$; t=0;
\State Compute Riemannian gradient $\mathbf{v}_{\mathrm{R}}^{0}$ according to Eq. (\ref{rmgradient}); Let $\bm{\mu}^{0}=-\mathbf{v}_{\mathrm{R}}^{0}$;
\Repeat
 \State Set $m^{t}=0$;
\While{$m^{t} \leq m_{\mathrm{max}}$}
    \If {Eq. (\ref{judge}) is true}
        \State \textbf{break};
    \EndIf
    \State $m^{t}=m^{t}+1$;
\EndWhile
\State Compute Armijo step size $\beta^{t}$ according to Eq. (\ref{Astep});
 \State Update $\mathbf{\bar{G}}^{t+1}$ according to Eq. (\ref{retractionop});
 \State Compute Riemannian gradient $\mathbf{v}_{\mathrm{R}}^{t+1}$ according to Eq. (\ref{rmgradient});
 \State Compute the vector transport result $\mathcal{T}_{\mathbf{\bar{G}}^{t} \to \mathbf{\bar{G}}^{t+1}} (\bm{\mu}^{t})$ according to Eq. (\ref{transportoperation});
 \State Compute Polak-Ribiere parameter $\alpha^{t+1}$ according to Eq. (\ref{PolakRibiere});
 \State Compute the update direction $\bm{\mu}^{t+1}$ according to Eq. (\ref{updatedirection});
 \State t=t+1;
 \Until{$\|\mathbf{v}_{\mathrm{R}}^{t}\|\leq \epsilon$}
\State  Recover $\{\{\mathbf{g}_{i,n}\}_{n=1}^{N}\}_{i=1}^{I}$ from
$\mathbf{\bar{G}}^{t}$  according to Eq. (\ref{defineG}) and Eq. (\ref{defineGi});
\State  Recover $\{\{\{\vartheta^{n}_{i,\bar{l}}\}_{n=1}^{N}\}_{\bar{l}=1}^{\bar{L}_{i}}\}_{i=1}^{I}$ from $\{\{\mathbf{g}_{i,n}\}_{n=1}^{N}\}_{i=1}^{I}$  according to
Eq. (\ref{recovephasefinal}).
\end{algorithmic}
\end{algorithm}

Let $\mathbf{\bar{G}}^{t}$ denote the updated variable at the $t$-th iteration. The tangent space of the complex circle manifold $\mathcal{O}$ at  $\mathbf{\bar{G}}^{t}$ consists of all the tangent matrices at $\mathbf{\bar{G}}_{t}$, denoted as $T_{\mathbf{\bar{G}}^{t}} \mathcal{O}$.  We have $T_{\mathbf{\bar{G}}^{t}} \mathcal{O}=\left\{\mathbf{X}\in \mathbb{C}^{(\sum_{i=1}^{I}\bar{L}_{i})\times N} \big{|} \Re\left\{\mathbf{X}\odot (\mathbf{\bar{G}}^{t})^{*}\right\}=\mathbf{0}\right\}$ \cite{absil2009optimization}. Among all the tangent vectors, the one that leads to the fastest increase in the objective function is defined as the Riemannian gradient, denoted as $\mathbf{v}^{t}_{\mathrm{R}}$. This gradient is calculated by projecting the Euclidean gradient onto $T_{\mathbf{\bar{G}}^{t}} \mathcal{O}$. Let $\mathbf{v}^{t}_{\mathrm{E}}$ denote the Euclidean gradient at $\mathbf{\bar{G}}^{t}$ and $f$ denote the objective function in problem $\mathcal{P}_{3}$. The Riemannian gradient $\mathbf{v}^{t}_{\mathrm{R}}$ is given by
\begin{align}\label{rmgradient}
\mathbf{v}^{t}_{\mathrm{R}}=\mathbf{v}^{t}_{\mathrm{E}}-\Re\left\{ \mathbf{v}^{t}_{\mathrm{E}} \odot (\mathbf{\bar{G}}^{t})^{*} \right\} \odot \mathbf{\bar{G}}^{t},
\end{align}
where $\mathbf{v}^{t}_{\mathrm{E}}$ is computed as
\begin{align}
\mathbf{v}^{t}_{\mathrm{E}}=\nabla_{\mathbf{\bar{G}}^{t}} f=-2\sum_{i=1}^{I}|\bar{\delta}_{i}|^2\mathbf{H}^{H}_{i}\pmb{\kappa}^{H}_{i}\mathbf{R}^{H}_{i}\mathbf{A}^{-2}\mathbf{R}_{i}\pmb{\kappa}_{i}\mathbf{H}_{i}\mathbf{\bar{G}}^{t},
\end{align}
with
\begin{align} \mathbf{A}=\sum_{i=1}^{I}\Re\{|\bar{\delta}_{i}|^2\mathbf{R}_{i}\pmb{\kappa}_{i}\mathbf{H}_{i}\mathbf{\bar{G}}\mathbf{\bar{G}}^{H}\mathbf{H}^{H}_{i}\pmb{\kappa}^{H}_{i}\mathbf{R}^{H}_{i}\}.
\end{align}

After the Riemannian gradient is obtained,  the variable in the Riemannian manifold is updated using Riemannian conjugate gradient \cite{absil2009optimization}, which is similar to  the Euclidean space in where the update direction at point $\mathbf{\bar{G}}^{t}$, denoted as $\bm{\mu}^{t}$, is given by $\bm{\mu}^{t} = -\mathbf{v}^{t}_{\mathrm{E}}+ \alpha^{t} \bm{\mu}^{t-1}$  \cite{avriel2003nonlinear}. Here, $\alpha^{t}$ is the Polak-Ribiere parameter, chosen to accelerate convergence. It indicates that the update direction is not only related to the updated gradient, but only related to the previous direction. However, in the Riemannian space, $\bm{\mu}^{t-1}$ and $\bm{\mu}^{t}$ are elements of $T_{\mathbf{\bar{G}}_{k-1}} \mathcal{O}$ and
$T_{\mathbf{\bar{G}}^{t}} \mathcal{O}$, respectively, and they should not be directly integrated across different tangent spaces. To address this, a vector transport operation \cite{absil2009optimization}  is introduced, which maps $\bm{\mu}^{t-1}$ in  $T_{\mathbf{\bar{G}}^{t-1}} \mathcal{O}$ to $T_{\mathbf{\bar{G}}^{t}} \mathcal{O}$. Specifically,
\begin{align}\label{transportoperation}
\mathcal{T}_{\mathbf{\bar{G}}^{t-1} \to \mathbf{\bar{G}}^{t}} (\bm{\mu}^{t-1})\triangleq \bm{\mu}^{t-1}- \Re\left\{ \bm{\mu}^{t-1} \odot (\mathbf{\bar{G}}^{t})^{*} \right\} \odot \mathbf{\bar{G}}^{t}.
\end{align}
Then, the update direction at $\mathbf{\bar{G}}^{t}$ is given by
\begin{align}\label{updatedirection}
\bm{\mu}^{t}=-\mathbf{v}_{\mathrm{R}}^{t}+\alpha^{t} \mathcal{T}_{\mathbf{\bar{G}}^{t-1} \to \mathbf{\bar{G}}^{t}} (\bm{\mu}^{t-1}).
\end{align}
Moreover, the Polak-Ribiere parameter is given by
\begin{align}\label{PolakRibiere}
\alpha^{t}=\frac{\mathrm{Tr}\left\{\Re\left\{\left(\mathbf{v}^{t}_{\mathrm{R}}\right)^{H} \left( \mathbf{v}^{t}_{\mathrm{R}}-\mathcal{T}_{\mathbf{\bar{G}}^{t-1} \to \mathbf{\bar{G}}^{t}} (\mathbf{v}_{\mathrm{R}}^{t-1}) \right) \right\} \right\}}{ \mathrm{Tr}\left\{\left(\mathbf{v}_{\mathrm{R}}^{t-1}\right)^{H}\mathbf{v}_{\mathrm{R}}^{t-1} \right\}}.
\end{align}

After obtaining the derived update direction, we can update $\mathbf{\bar{G}}^{t+1}$ as $\mathbf{\bar{G}}^{t}+\beta^{t}\bm{\mu}^{t}$. $\beta^{t}$ is the Armijo step size according to the line search algorithm  \cite{avriel2003nonlinear}, which is widely used to accelerate  convergence. However, in this way, the updated variable may lie outside the  complex circle manifold. Therefore, a retraction operation \cite{absil2009optimization} is used to update the variable as follows.
\begin{align}\label{retractionop}
\mathbf{\bar{G}}^{t+1}=\mathcal{R}_{\mathbf{\bar{G}}^{t}}\left(\beta^{t}\bm{\mu}^{t}\right)\triangleq
\mathrm{unit}\left\{\mathbf{\bar{G}}^{t}+\beta^{t}\bm{\mu}^{t}\right\}.
\end{align}
One can readily verify that $\mathbf{\bar{G}}^{t+1}$ in Eq. (\ref{retractionop}) is on the complex circle manifold. Moreover, given scalars $\alpha_{A}> 0$, $\beta_{A}\in(0, 1)$ and $\sigma\in(0, 1)$, the Armijo step size $\beta^{t}$ is calculated by
\begin{align}\label{Astep}
\beta^{t}=\alpha_{A}\beta_{A}^{m^{t}},
\end{align}
where $m^{t}$ is the smallest nonnegative integer such that
\begin{align}\label{judge}
&f\left(\mathcal{R}_{\mathbf{\bar{G}}^{t}}\left(\alpha_{A}\beta_{A}^{m^{t}}\bm{\mu}^{t}\right)\right)-f\left(\mathbf{\bar{G}}^{t}\right)\\\nonumber
&\leq\sigma\alpha_{A}\beta^{m^{t}}_{A}\Re\left\{\mathrm{Tr}\left\{(\bm{\mu}^{t})^{H}\mathbf{v}^{t}_{\mathrm{R}} \right\}\right\}
\end{align}

Since the objective function is monotonically non-increasing in each iteration, Algorithm 1 is guaranteed to converge to a stationary point of the smooth objective function in problem  $\mathcal{P}_{3}$ \cite{avriel2003nonlinear}.

To show the computing complexity of Algorithm 1, we provide the floating-point operations (FLOPs) of it. For clarity and simplicity, we give the expression by ignoring all terms except for the leading (highest order or dominant) terms and let $\bar{L}_{i}=\bar{L},~\forall i$. Then the worst FLOPs of Algorithm 1, denoted as $\mathcal{O}_{CPSOA-RM}$, is given by
\begin{align}
&\mathcal{O}_{CPSOA-RM}\\\nonumber
&=\mathcal{O}\left(N_{RM}I^3NL^2+N_{RM}m_{max}\left(I^2NL+IN^2L\right)\right),
\end{align}
where $N_{RM}$ is the  number of iterations of Algorithm 1.

\subsection{Discrete Phase Shift Optimization}
In many scenarios, continuous phase shift is hard to achieve, but discrete phase shift is implementable \cite{chen2025integrated}.  Without loss of generality, we assume  $N_{B}$ bits are usable to control the phase shift of each RIS element by the controller (e.g., via PIN diodes) \cite{di2020hybrid}. Thus there are $\ddot{B}$ possible phase shifts of each RIS element with $\ddot{B}=2^{N_{B}}$. The optimization problem on discrete phase shifts can be formed  as follows.
\begin{align}\label{pl}
\mathcal{P}_{4}: &\min_{\{\{\{\vartheta^{n}_{i,\bar{l}}\}_{n=1}^{N}\}_{\bar{l}=1}^{\bar{L}_{i}}\}_{i=1}^{I}}~~~~~~~~~~\mathrm{L_R}    \nonumber\\
&~~~~~~~\text{s. t.}~~~~~~~~~~~~~~~|\mathbf{g}_{i,n}(\bar{l})|=1,\\\nonumber
&~~~~~~~~~~~~~~~~~~~~~~~~~\vartheta^{n}_{i,\bar{l}}=\angle\left\{\mathbf{g}_{i,n}(\bar{l})\right\},\\\nonumber
&~~~~~~~~~~~~~~~~\vartheta^{n}_{i,\bar{l}}\in\left\{c_{b}\frac{2\pi}{\ddot{B}}\big|c_{b}\in\mathbb{Z},c_{b}\leq \ddot{B}\right\},\\\nonumber
&~~~~n=1,2,\cdots,N, ~\bar{l}=1,2,\cdots,\bar{L}_{i}, ~i=1,2,\cdots,I.
\end{align}

Problem $\mathcal{P}_{4}$ is challenging to address with the non-linear objective function, the high-dimensional matrix variable, unit-modulus constrains and discrete phase shift constrains. This optimization problem presents inherent complexities that defy standard convex approximation techniques and gradient-based optimization routines. We consequently devise a swarm intelligence algorithm named DPSOA-I-GWO to solve it, as outlined in Algorithm \ref{alg_2}. The proposed DPSOA-I-GWO is developed on the improved grey wolf optimizer (I-GWO) \cite{nadimi2021improved}. I-GWO is recently proposed based on the leadership hierarchy and group hunting mechanism of the grey wolves \cite{nadimi2021improved}, and widely used in engineering problems. However, I-GWO cannot solve $\mathcal{P}_{4}$ directly, since it only solves continuous optimization problems.

\begin{algorithm}
\caption{Discrete Phase Shift Optimization Algorithm Based on I-GWO (DPSOA-I-GWO)}
\label{alg_2}
\begin{algorithmic}[1]
\Require Number of iterations $T$;  Number of wolves $M$.
\Ensure  Phase shifts of RISs $\{\{\{\vartheta^{n}_{i,\bar{l}}\}_{n=1}^{N}\}_{\bar{l}=1}^{\bar{L}_{i}}\}_{i=1}^{I}$.
\State \textbf{Initialize}: Randomly distribute $M$ wolves in the $D$-dimensional variable space by Eq.~(\ref{inige}),  and calculate their fitness values by Eq.~(\ref{fitnesseq}); t=0;
\Repeat
\State t=t+1;
\State Find the first three best wolves $\alpha$, $\beta$ and $\gamma$ as leader wolves;
\For {$m=1$ to $M$}
\State  Compute  $\mathbf{X}_{m}^{\alpha}(t)$, $\mathbf{X}_{m}^{\beta}(t)$ and $\mathbf{X}_{m}^{\gamma}(t)$ by Eq.~(\ref{grheq});
\State Compute the first candidate learned from leader wolves by  Eq.~(\ref{grhfeq});
\State Construct neighborhoods of  the $m$-th wolf by Eq.~(\ref{neieq});
\For {$d=1$ to $D$}
\State Compute the $d$-th dimension of the second candidate learned from neighborhoods by  Eq.~(\ref{leabyneieq});
\EndFor
\State Select the best candidate by Eq.~(\ref{selecteq});
\State Update the $m$-th wolf to $\mathbf{X}_{m}(t)$ by Eq.~(\ref{updateeq});
\EndFor
\Until {t=T}
\State Compute the best wolf position  $\mathbf{X}^{*}$ by Eq.~(\ref{bestwolfeq});
\State Recover the phase shifts of RISs  $\{\{\{\vartheta^{n}_{i,\bar{l}}\}_{n=1}^{N}\}_{\bar{l}=1}^{\bar{L}_{i}}\}_{i=1}^{I}$ from $\mathbf{X}^{*}$   by Eq.~(\ref{finaltrsform}).
\end{algorithmic}
\end{algorithm}

DPSOA-I-GWO starts by an initializing. In this phase, $M$ wolves are randomly distributed in the limited variable space of problem $\mathcal{P}_{4}$. The position of the $m$-th wolf is denoted as $\mathbf{X}_{m}$ with $\mathbf{X}_{m}\in \mathbb{R}^{D\times1}$. The term $D$ is the dimensionality of variable in problem $\mathcal{P}_{4}$ , i.e, $ N\sum_{i=1}^{I}\bar{L}_{i}$.  The $d$-th dimension of $\mathbf{X}_{m}$ is given by

\begin{align}\label{inige}
X_{m}(d)=r_{m,d}\frac{2\pi}{\ddot{B}},
\end{align}
where $r_{m,d}=\mathrm{randi}[0,\ddot{B}]$, which is a random integer in the range $[0,\ddot{B}]$.

Then the fitness values of wolves denoted as $f_{\mathrm{GWO}}$ are calculated.  For the $m$-th wolf, the fitness value is given by
\begin{align}\label{fitnesseq}
f_{\mathrm{GWO}}(\mathbf{X}_{m})=\mathrm{L_R}\left(\{\{\{\vartheta^{n}_{i,\bar{l}}\}_{n=1}^{N}\}_{\bar{l}=1}^{\bar{L}_{i}}\}_{i=1}^{I}\right),
\end{align}
where
\begin{align}\label{gandx}
\vartheta^{n}_{i,\bar{l}}=\mathbf{X}_{m}\left(N\sum_{i^{'}=1}^{i-1}\bar{L}_{i^{'}}+(n-1)\bar{L}_{i}+\bar{l}\right).
\end{align}

Then the following steps including movement and selecting and updating are repeated until the stopping criterion is satisfied. The stopping criterion of DPSOA-I-GWO  is to reach the predefined number of iterations denoted as $T$. Let $t$ represent the current number of iterations.

The movement phase involves the group hunting and the individual hunting. In this phase, the first three best wolves from the wolf pack are selected as leader wolves named $\alpha$, $\beta$ and $\gamma$ with positions denoted as $\mathbf{X}^{\alpha,t}$, $\mathbf{X}^{\beta,t}$ and $\mathbf{X}^{\gamma,t}$, respectively. After that, the linearly decreased coefficient $a_{G}^{t}$ , and coefficients $\{A_{k,d}^{t}\}_{k=1}^{3}$ and $\{C_{k,d}^{t}\}_{k=1}^{3}$ are calculated according to the follow equations.
\begin{subequations}
\begin{align}
a_{G}^{t}=2-2\left(\frac{t}{T}\right)^2,
\end{align}
\begin{align}
A_{k,d}^{t}=(2\mathrm{rand}[0,1]-1)a_{G}^{t},
\end{align}
\begin{align}
C_{k,d}^{t}=2\mathrm{rand}[0,1],
\end{align}
\end{subequations}
where $\mathrm{rand}$ represents the operation of generating a random number. The group hunting led by the three leader wolves is considered in the movement phase. In this way, the prey encircling is determined considering  $\mathbf{X}^{\alpha,t}$, $\mathbf{X}^{\beta,t}$ and $\mathbf{X}^{\gamma,t}$ by the following equations.
\begin{subequations}\label{grheq}
\begin{align}
\mathbf{X}_{m}^{\alpha,t}(d)=\mathbf{X}^{\alpha,t}(d)-A_{1,d}^{t}|C_{1,d}^{t}\mathbf{X}^{\alpha,t}(d)-\mathbf{X}^{t}_{m}(d)|,
\end{align}
\begin{align}
\mathbf{X}_{m}^{\beta,t}(d)=\mathbf{X}^{\beta,t}(d)-A_{2,d}^{t}|C_{2,d}^{t}\mathbf{X}^{\beta,t}(d)-\mathbf{X}^{t}_{m}(d)|,
\end{align}
\begin{align}
\mathbf{X}_{m}^{\gamma,t}(d)=\mathbf{X}^{\gamma,t}(d)-A_{3,d}^{t}|C_{3,d}^{t}\mathbf{X}^{\gamma,t}(d)-\mathbf{X}^{t}_{m}(d)|.
\end{align}
\end{subequations}
Then, the first candidate for the new position of the $m$-th wolf named $\mathbf{\dot{X}}^{t+1}_{m}$ is calculated by the following equation.
\begin{align}\label{grhfeq}
\mathbf{\dot{X}}^{t+1}_{m}(d)=\mathcal{T}_{\mathrm{Bit}}\left\{\frac{1}{3}\mathbf{X}_{m}^{\alpha,t}\big(d\big)+ \frac{1}{3}\mathbf{X}_{m}^{\beta,t}\big(d\big)+\frac{1}{3}\mathbf{X}_{m}^{\gamma,t}\big(d\big)\right\},
\end{align}
where $\mathcal{T}_{\mathrm{Bit}}$ is a retraction function aiming to meet the quantization constraint of problem $\mathcal{P}_{4}$, which is given by
\begin{align}
\mathcal{T}_{\mathrm{Bit}}\left\{\mathbf{x}\right\}=\frac{2\pi}{\ddot{B}}\mathrm{round}\left\{\mathrm{mod}\left\{\frac{\ddot{B}x}{2\pi},\ddot{B}\right\} \right\},~ \forall x\in \mathcal{R},
\end{align}
where $\mathrm{round}$ represents the rounding operation and $\mathrm{mod}$ represents the modulo operation.

Besides the group hunting, the individual hunting learned by neighbors is also considered in the movement phase. For doing this, a radius $R^{t}_m$ is calculated by $R^{t}_m=\|\mathbf{X}^{t}_{m}- \mathbf{\dot{X}}^{t+1}_{m} \|$. Then the neighbors of $\mathbf{X}^{t}_{m}$ denoted by $\mathbf{N}^{t}_{m}$ is constructed by
\begin{align}\label{neieq}
&\mathbf{N}^{t}_{m}=\{\mathbf{X}^{t}_{n}\big| \| \mathbf{X}^{t}_{m}-\mathbf{X}^{t}_{n}\|\leq R^{t}_m,~  n=1,2,\cdots,M\}.
\end{align}
Then the learning by neighbors is performed. The learning result of the $m$-th wolf denoted as $\mathbf{\ddot{{X}}}^{t+1}_{m}$ is the second candidate,  and the $d$-th dimension of it is expressed as
\begin{align}\label{leabyneieq}
&\mathbf{\ddot{{X}}}^{t+1}_{m}(d)=\mathcal{T}_{\mathrm{Bit}}\left\{\mathbf{X}^{t}_{m}(d)+\mathrm{rand}[0,1](\mathbf{X}^{'}_{m}(d)-\mathbf{X}_{r}(d))\right\},
\end{align}
where $\mathbf{X}^{'}_{m}$ is  a random element in $\mathbf{N}^{t}_{m}$, $\mathbf{X}_{r}$ is a random wolf  from the wolf pack.

In the selecting and updating phase, the superior candidate from the two candidates is selected as follows.
\begin{align}\label{selecteq}
&\mathbf{\grave{X}}_{m}^{t+1}=\begin{cases}
\mathbf{\dot{X}}^{t+1}_{m}, & f_{\mathrm{GWO}}(\mathbf{\dot{X}}^{t+1}_{m})<f_{\mathrm{GWO}}(\mathbf{\ddot{{X}}}^{t+1}_{m})   \\
\mathbf{\ddot{{X}}}^{t+1}_{m}, &  f_{\mathrm{GWO}}(\mathbf{\dot{X}}^{t+1}_{m})\geq f_{\mathrm{GWO}}(\mathbf{\ddot{{X}}}^{t+1}_{m})
\end{cases}.
\end{align}
Then the original position of wolf is also compared with the superior candidate to update the new position, shown as follows.
\begin{align}\label{updateeq}
&\mathbf{X}^{t+1}_{m}=\begin{cases}
\mathbf{X}^{t}_{m}, & f_{\mathrm{GWO}}(\mathbf{X}^{t}_{m})<f_{\mathrm{GWO}}(\mathbf{\grave{X}}^{t+1}_{m})   \\
\mathbf{\grave{X}}^{t=1}_{m}, &  f_{\mathrm{GWO}}(\mathbf{X}^{t}_{m})\geq f_{\mathrm{GWO}}(\mathbf{\grave{X}}^{t+1}_{m})
\end{cases}.
\end{align}

After updating for all individuals, the new round of iteration comes. When the predefined number of iterations is reached, the best position of the prey denoted as $\mathbf{X}^{*}$ is located as follows.
\begin{align}\label{bestwolfeq}
\mathbf{X}^{*}=\mathop{\arg\min}\limits_{\mathbf{X}_{m}\in \{ \mathbf{X}^{T}_{m}\}_{m=1}^{M}} f_{\mathrm{GWO}}(\mathbf{X}_{m}).
\end{align}
Finally,  the optimized phase shifts are obtained as follows.
\begin{align}\label{finaltrsform}
\vartheta^{n}_{i,\bar{l}}=\mathbf{X}^{*}(N\sum_{i^{'}=1}^{i-1}\bar{L}_{i^{'}}+(n-1)\bar{L}_{i}+\bar{l}).
\end{align}

Here, we provide the computing complexity of Algorithm~\ref{alg_2} by FLOPs. Similar to Algorithm~\ref{alg_1}, we assume $\bar{L}_{i}=\bar{L},~\forall i$ and ignore all terms except for the leading terms. Specifically, the FLOPs of Algorithm~\ref{alg_2}, denoted as  $\mathcal{O}_{\mathrm{DPSOA-I-GWO}}$, is expressed as
\begin{align}
\mathcal{O}_{\mathrm{DPSOA-I-GWO}}=\mathcal{O}\left(N_{\mathrm{I-GWO}}\left(NIM^2+INLM\right)\right),
\end{align}
where $N_{\mathrm{I-GWO}}$ is the the number of iterations of DPSOA-I-GWO.

\section{Simulation Results}
In this section, we provide simulation results to discuss the system performance of the RISs-assisted mmWave position sensing and demonstrate the effectiveness of proposed phase shift optimization algorithms.

\subsection{Settings of Numerical Experiment}
The RISs-assisted mmWave position sensing system is deployed in an indoor environment in which the RISs are installed on the wall.  Without loss of general, the size of the indoor environment is  $8~\mathrm{m} \times 5~\mathrm{m} \times 4~\mathrm{m}$,  the AP is located at $[5,1,0]~(\mathrm{m})$ equipped with an ULA, and its array direction is $[0,1,0]$. The UE is located at $[4,4,1]~(\mathrm{m})$ with a single antenna. By default, two RISs on the wall, named RIS 1 and RIS 2, are available to generate VLoS paths. The positions of RIS 1 is $[0,3,2]~(\mathrm{m})$, and the orientation is given by $\mathbf{e}_{x,1}=[0,-1,0]$, $\mathbf{e}_{y,1}=[0,0,1]$ and $\mathbf{e}_{z,1}=[1,0,0]$. The positions of RIS 2 is $[8,3,3]~(\mathrm{m})$, and the orientation is given by $\mathbf{e}_{x,2}=[0,-1,0]$, $\mathbf{e}_{y,2}=[0,0,1]$ and $\mathbf{e}_{z,2}=[-1,0,0]$.

The operation frequency of this system is about $28~\mathrm{GHz}$ with wavelength $\lambda=0.0108~\mathrm{m}$. By default, the transmit power of AP  $P_{0}=5~\mathrm{dBm}$, the noise  power $\sigma_{w}^2=-90~\mathrm{dBm}$, and the number of measurements $N=50$. There are $5$ non-negligible NLoS paths and each path coefficient follows complex Gaussian distribution, i.e., $\delta_{l}\in \mathcal{N}(0,\sigma_{l}^2)$ with $10\lg{\frac{|\delta_{1}|^2}{\sigma_{l}^2}}=40~\mathrm{dB},~l=1,2,\cdots,5$. Note that $\delta_{1}$ is the channel gain of the LoS path. And the AoD at AP of the NLoS path follows uniform distribution, i.e, $\theta_{AU,l}\in\mathcal{U}(0,\pi]$.
The default parameters about the size of RISs are set as follows: $d_{x_{i}}=d_{y_{i}}=0.01~\mathrm{m},N_{I,i}=8,~M_{I,i}=8,~\forall i$. The default parameters about the size of AP are set as follows: $N_{A}=16,~\Delta d=0.0054~\mathrm{m}$. The antenna gain of AP $G_{t}=8~\mathrm{dBi}$ while the antenna gain of UE $G_{r}=0~\mathrm{dBi}$. The amplitude gain of
reflective element $A_{i}=1,~\forall i$.

\subsection{Experimental Results on Continuous Phase Shifts}
In this subsection, we present experimental results pertaining to the position sensing system employing RISs with continuous phase shifts, along with the performance evaluation of the proposed CPSOA-RM outlined in Algorithm \ref{alg_1}. Based on experimental trials, the parameters of CPSOA-RM are configured as follows: $\epsilon=10^{-6}$, $m_{\mathrm{max}}=200$, $a_{A}=10^{5}$ $\beta_{A}=0.5$, $\sigma=0.1$. To verify the effectiveness of it, the EBS algorithm in \cite{alhafid2024enhanced} is adopted as the benchmark where the sweeping directions  are uniform distributed within the measurements for per VLoS path.

\begin{figure}
   \subfloat[ \label{PEBTPs1}$N=50$.]{%
       \includegraphics[width=0.45\textwidth]{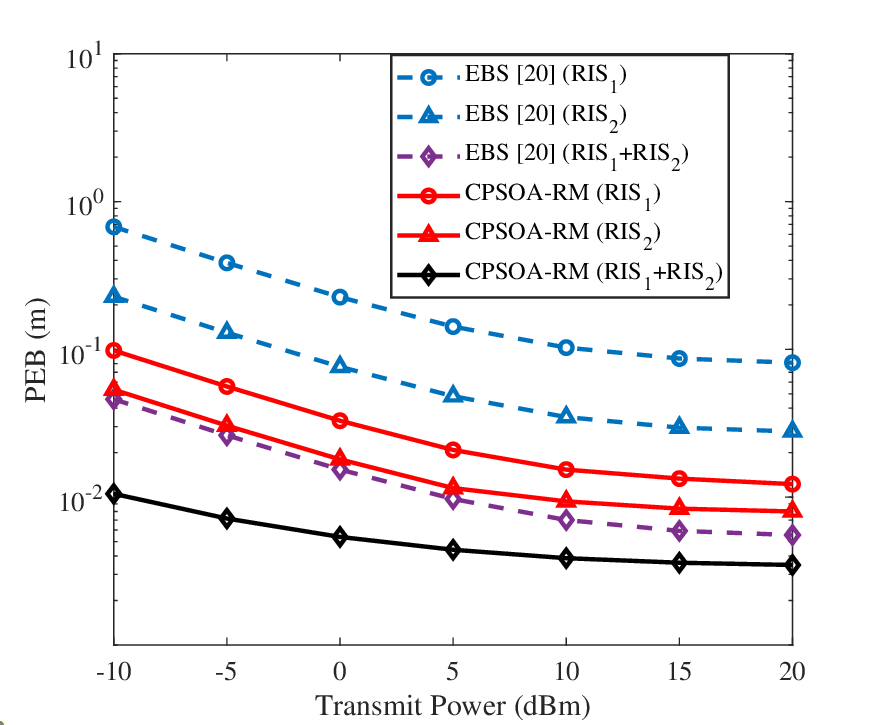}
     }
     \\\vspace{-0.5\baselineskip}
      \subfloat[ \label{PEBTPs2}$N=10$.]{%
       \includegraphics[width=0.45\textwidth]{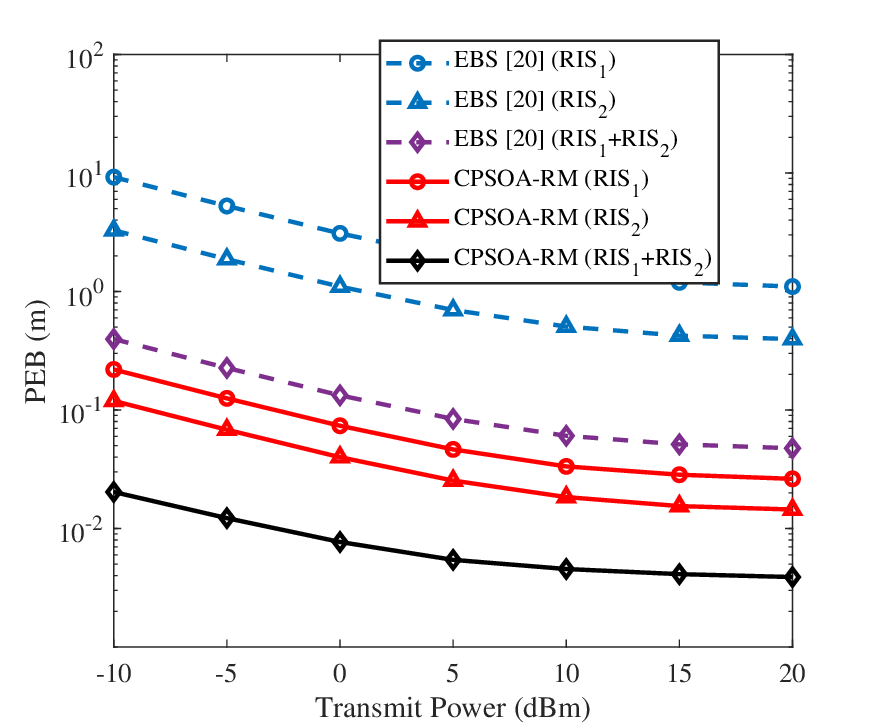}
     }
     \caption{PEB versus transmit power.}
     \label{PEBTP}
     \vspace{-1\baselineskip}   
\end{figure}

Fig.~\ref{PEBTP} plots the PEBs versus the transmit power with different measurement amounts. It shows that the proposed CPSOA-RM consistently outperforms EBS  in terms of PEB across varying transmit power levels. And the PEB with two RISs is much smaller than the single RIS.  A notable point is that PEB based on CPSOA-RM is lower than $0.01~\mathrm{m}$ when the transmit power is lager than $-5$ dBm for $N=50$ and lager than $0$ dBm for $N=10$. Therefore, a sub-centimeter positioning accuracy is easily achieved by using the proposed CPSOA-RM.

As observed from Fig.~\ref{PEBTP}, the PEBs for both CPSOA-RM and EBS generally decrease with increasing transmit power. When at the low transmit power region, the decreasing trends of PEBs are obvious, but at the high power region beyond 5 dBm, the PEBs almost converge. This is because the  increasing transmit power not only enhances the signal strength of the VLoS path, but also enhance that of the NLoS path, which is non-negligible in mmWave position sensing. As the transmit power increases, NLoS paths gradually dominate the interference,  the system transitions from a noise limited region to an interference limited region. When comparing Fig.~\ref{PEBTP}\subref{PEBTPs1} and Fig.~\ref{PEBTP}\subref{PEBTPs2}, we can conclude that a larger measurement amount results in a smaller PEB for the proposed CPSOA-RM or EBS. Furthermore, conducting more measurements yields significant improvement when the system is noise-limited (e.g., in single-RIS scenarios or at low transmit power regions), but provides little benefit when the system is interference-limited.


\begin{figure}
  \centering
  \includegraphics[width=0.45\textwidth]{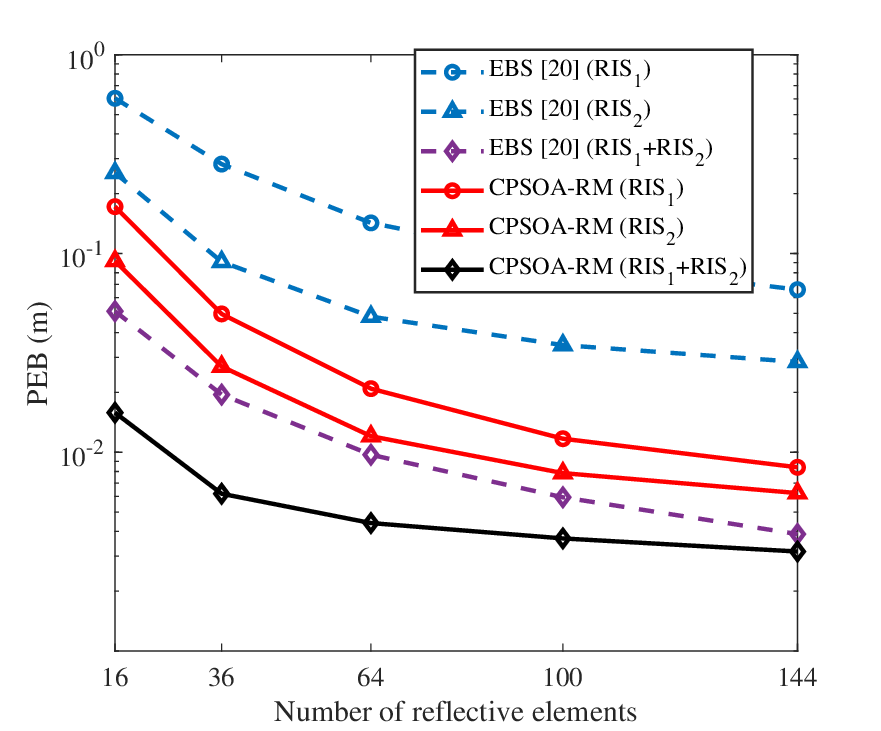}\\
  \caption{PEB versus number of reflective elements.}\label{PEBNRE}
  \vspace{-1\baselineskip}   
\end{figure}

Fig.~\ref{PEBNRE} plots the PEBs versus the number of reflective elements. As seen, the PEBs based on CPSOA-RM and EBS both decrease with the growth of number of reflective elements. This is attributed to the fact that with more reflective elements, the signal strength as well as the sensing DoFs of VLoS path will increase, which raise the positioning information quantity. Moreover, increasing the number of reflective elements narrows the beamwidth of the reflected beam of RIS. As a result, more NLoS components are effectively suppressed, and the dominant VLoS path is further strengthened, which leads to the PEB reduction. From Fig.~\ref{PEBNRE}, we conclude that the proposed CPSOA-RM is superior to EBS in terms of  PEB across different numbers of reflective elements.

\begin{figure}
  \centering
  \includegraphics[width=0.45\textwidth]{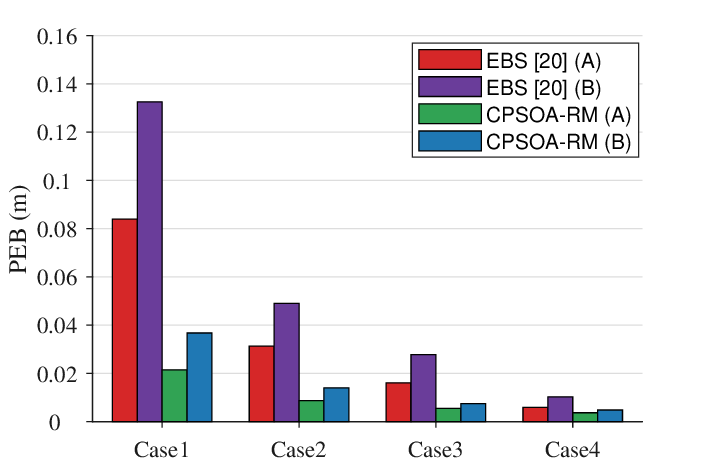}\\
  \caption{PEBs under different multiple-RISs scenarios (In case 1, $P_{0}=-5$ dBm and $L_{I}=25$; In case 2, $P_{0}=5$ dBm and $L_{I}=25$; In case 3, $P_{0}=-5$ dBm and $L_{I}=100$; In case 4, $P_{0}=5$ dBm and $L_{I}=100$).}\label{PEBRISP}
  \vspace{-1\baselineskip}   
\end{figure}

Fig.~\ref{PEBRISP} shows the PEBs under two multiple-RIS scenarios with different RIS positions. In scenario A, RIS 1 and RIS 2 are both available. In scenario B, RIS 1 and RIS 3 are both available where RIS 3 is located at [0,2,3] (m) with the same size of RIS 2. Four cases are considered in Fig.~\ref{PEBRISP}, which are combinations of different powers and numbers of reflective elements, represented by $L_{I}$.  As observed from Fig.~\ref{PEBRISP}, the PEBs based on CPSOA-RM or EBS in scenario A is always lower than scenario B under different cases. This is caused by the discrepancy of geometry structure of RISs in the two scenarios. The two RISs in scenario A are installed on the opposite walls  respectively while the two RISs in scenario B are installed on the same wall. The geometry structure of RISs in scenario A is better than scenario B. Thus, the RIS position plays an important role on PEB.

\begin{figure}
   \subfloat[ \label{PEBRA1}One RIS.]{%
       \includegraphics[width=0.23\textwidth]{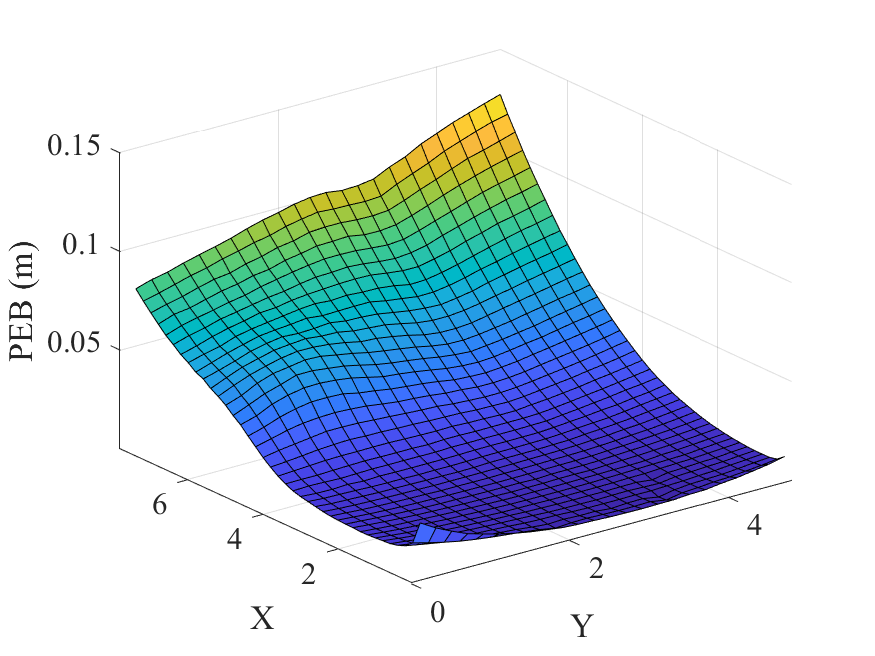}
     }\hfill
      \subfloat[ \label{PEBRA2}Two RISs.]{%
       \includegraphics[width=0.23\textwidth]{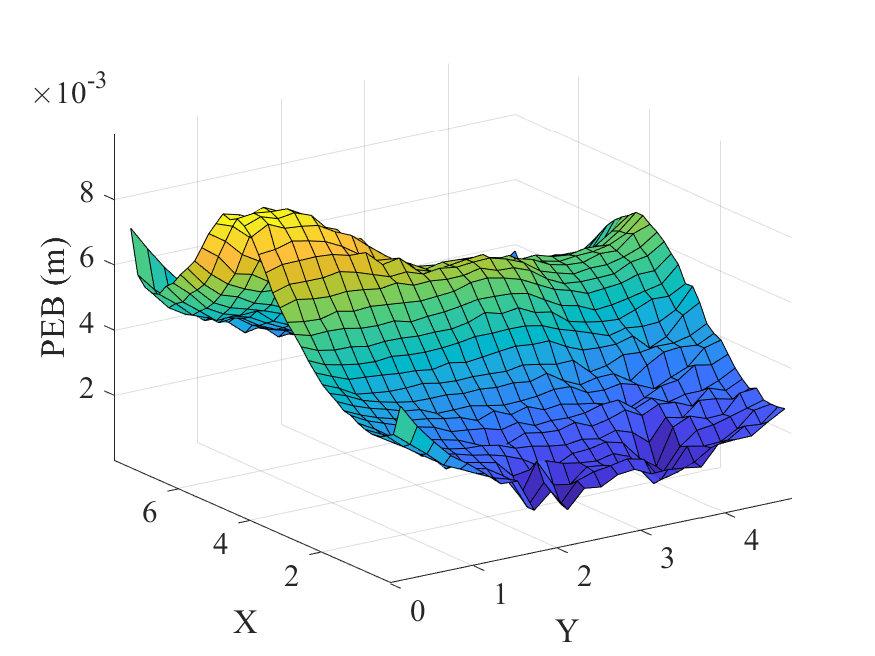}
     }\\
      \subfloat[ \label{PEBRA3}Three RISs.]{%
       \includegraphics[width=0.23\textwidth]{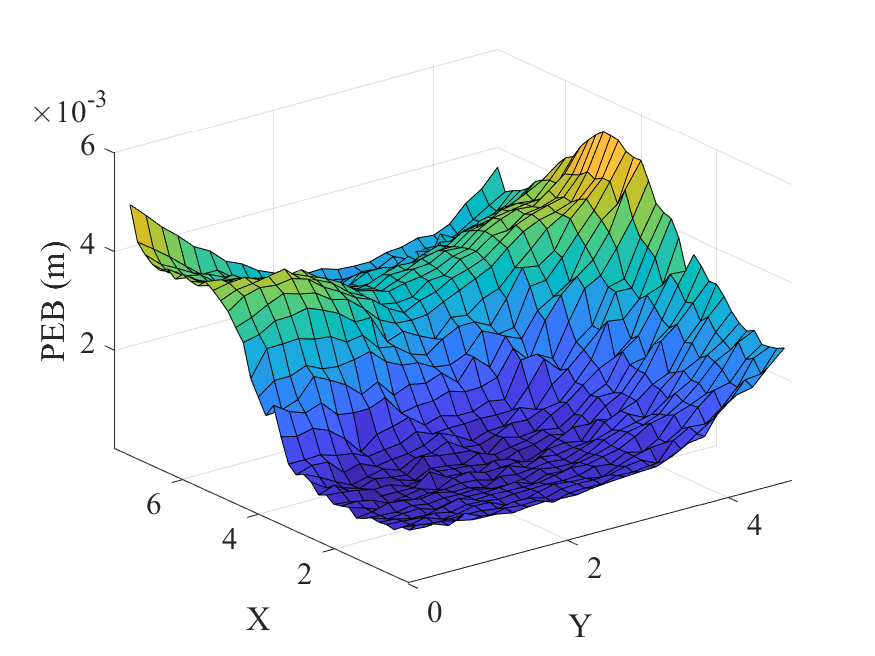}
     }\hfill
      \subfloat[ \label{PEBRA4}Four RISs.]{%
       \includegraphics[width=0.23\textwidth]{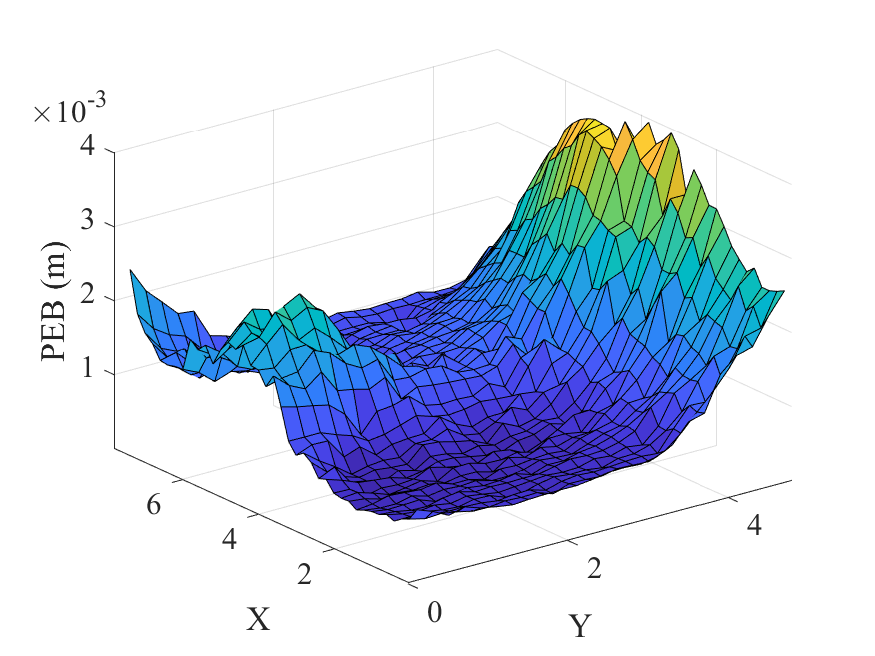}
     }
     \caption{PEBs on a plane with different RIS amounts.}
     \vspace{-1\baselineskip}   
     \label{PEBRA}
\end{figure}

Fig.~\ref{PEBRA} shows the PEBs based on the proposed CPSOA-RM versus the position of UE on a plane with different RIS amounts. The height of plane is 1 m. In Fig.~\ref{PEBRA}\subref{PEBRA1}, only RIS 1 is available.  In Fig.~\ref{PEBRA}\subref{PEBRA2}, RIS 1 and RIS 2 are available. Besides the two RISs, the extra third available RIS with position [0, 1, 3] (m) is available in Fig.~\ref{PEBRA}\subref{PEBRA3}. In Fig.~\ref{PEBRA}\subref{PEBRA4}, the fourth available RIS with position [8, 4, 2] (m) is involved. From these subfigures, it is observed that the PEB of UE always becomes smaller when the UE is closer to the RIS. As seen in Fig.~\ref{PEBRA}, when the RIS amount becomes large, the PEB of UE at the same position is decreased. And the improvement of positioning accuracy by employing an extra RIS is significant when the amount of existing RISs is small. Moreover,  the worst positioning accuracy in the interested area is improved when more RISs are available. Specially, the worst positioning accuracy is decreased from $0.12$ m  to $0.004$ m when RIS amount is increased from  one to four.

\begin{figure}
  \centering
  \includegraphics[width=0.45\textwidth]{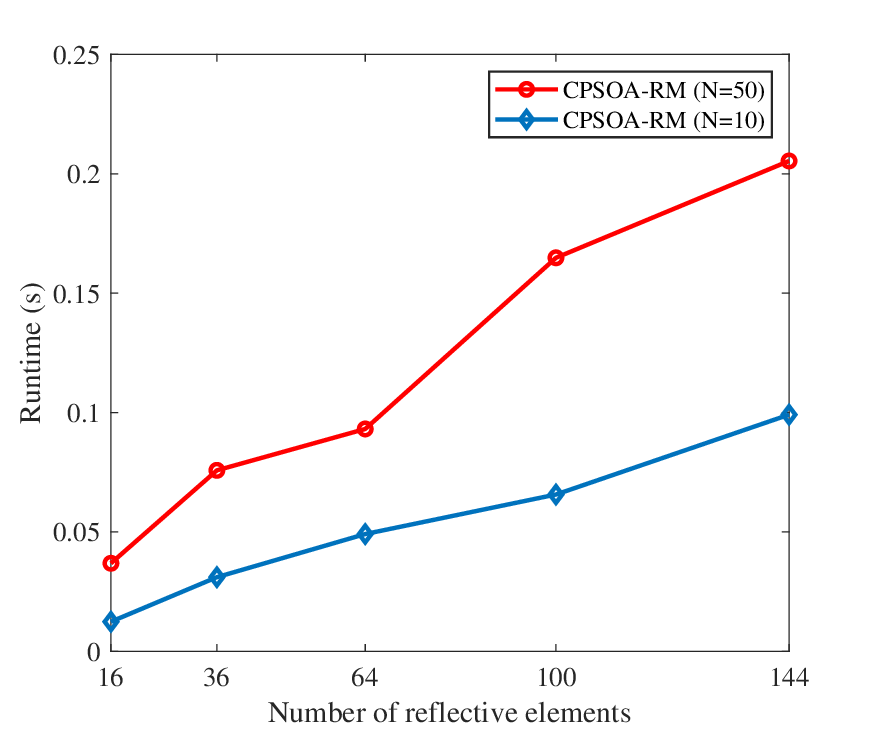}\\
  \caption{Runtime of CPSOA-RM versus number of reflective elements.}\label{rumtimeconfig}
  \vspace{-1\baselineskip}   
\end{figure}

Fig.~\ref{rumtimeconfig} shows the runtime of the proposed CPSOA-RM versus number of reflective elements. The simulation results are obtained on a host equipped with an Intel i9-14900HX processor. As expected, the the runtime increases as the number of reflective elements becomes larger.  Moreover, taking more measurements  also increases the runtime. This figure also indicates that CPSOA-RM is suitable for large-scale deployments. For example, when $L_{I}=144$ and $N=50$, the runtime of CPSOA-RM is only about 0.2 s.

\subsection{Experimental Results on Discrete Phase Shifts}
In this subsection, we present experimental results pertaining to the position sensing system  employing RISs with discrete phase shifts, along with the performance evaluation of the proposed DPSOA-I-GWO outlined in Algorithm \ref{alg_2}.  The default number of bits controlling the phase shift of RIS is set to be $2$, i.e., $N_{B}=2$. Therefore, there are four different states for the phase shift of each reflective element. After experimental trials, we set $T=1000$, and $M=100$ in Algorithm \ref{alg_2}. To demonstrate the effectiveness of it, a discrete scheme of the EBS proposed in \cite{alhafid2024enhanced} is chosen as the benchmark, named discrete EBS. Specially, in this scheme, the phase shift is obtained by modifying each continuous phase shift obtained by EBS to the nearest possible discrete sate.

\begin{figure}
   \subfloat[ \label{NDISDN1}$N=50$.]{%
       \includegraphics[width=0.45\textwidth]{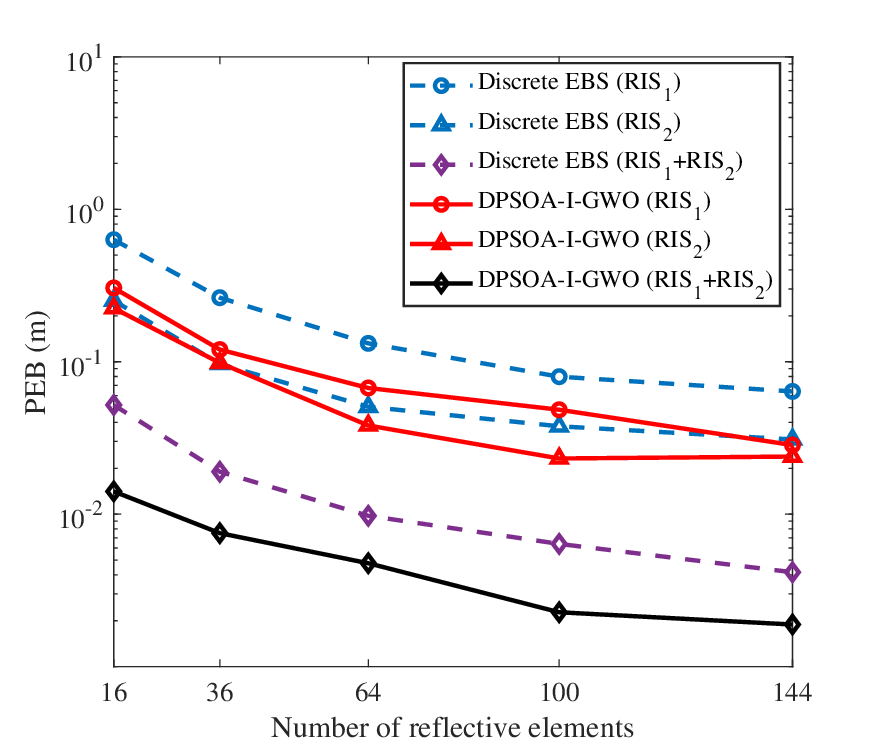}
     }
     \\\vspace{-0.5\baselineskip}
      \subfloat[ \label{NDISDN2}$N=25$.]{%
       \includegraphics[width=0.45\textwidth]{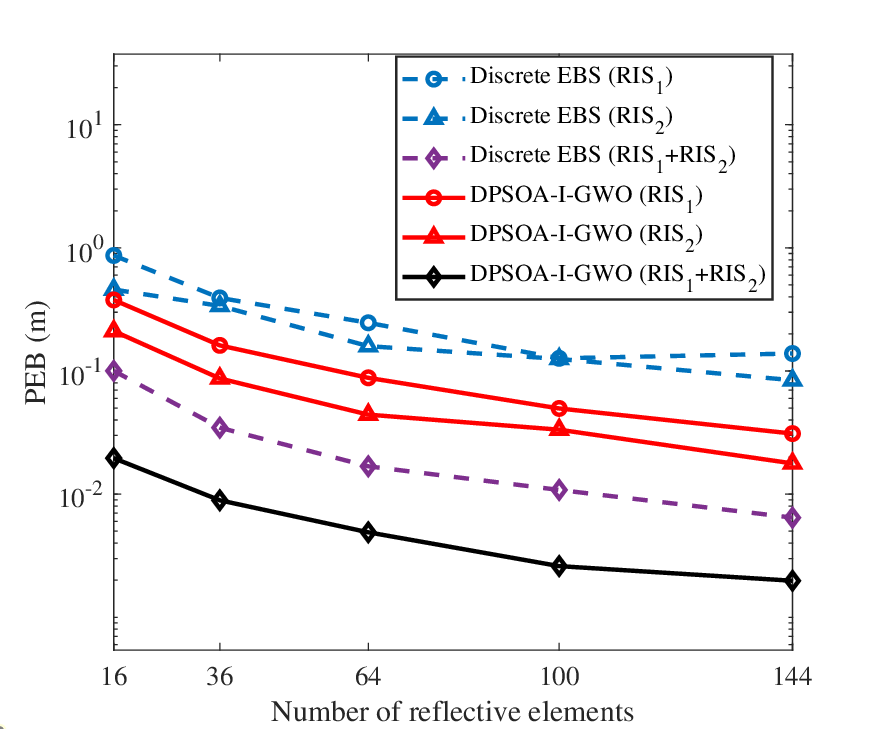}
     }
     \caption{PEB versus number of reflective elements in the discrete configuration.}
     \label{NDISDN}
     \vspace{-1\baselineskip}   
\end{figure}

Fig.~\ref{NDISDN} plots the PEBs versus the number of reflective elements.
As expected, the PEB decreases with the increasing number of reflective elements, especially at the low region. Since more DoFs are provided even with limited sates of phase shift when more reflective elements are involved. Moreover, the proposed DPSOA-I-GWO is obviously superior to the discrete EBS in terms of PEB across different numbers of reflective elements. For example, when $N=50$ and $L_{I}=100$, the PEB of discrete EBS is about 0.01 m, but the PEB of DPSOA-I-GWO is about 0.002 m in the two-RISs case. When comparing Fig.~\ref{NDISDN}\subref{NDISDN1} and Fig.~\ref{NDISDN}\subref{NDISDN2}, we can conclude that the PEB decreases with more measurements. However, the proposed DPSOA-I-GWO still performs well with low amount of measurements. For example, when $N=25$ and $L_{I}=64$, the PEB of DPSOA-I-GWO is about 0.004 m.

\begin{figure}
  \centering
  \includegraphics[width=0.45\textwidth]{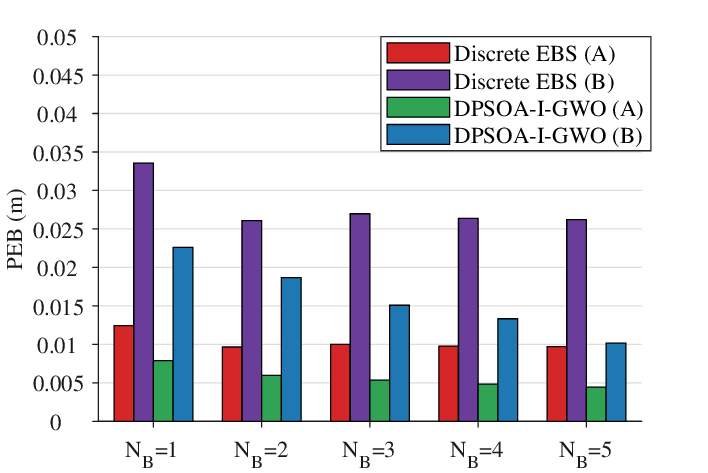}\\
  \caption{PEB versus number of bits.}\label{PEBBitsdis}
  \vspace{-1\baselineskip}   
\end{figure}

Fig.~\ref{PEBBitsdis} plots the PEBs with different numbers of bits under two scenarios. In scenario A, the transmit power is equal to $5$ dBm while equal to $-5$ dBm in scenario B. As seen in the figure, a larger transmit power leads to a lower PEB across numbers of bits. Moreover,
the proposed DPSOA-I-GWO is over  discrete EBS across numbers of bits. When the number of bits increases from one to two, the improvement of  discrete EBS is evident. But when the number of bits is larger than two, the PEB of discrete EBS is almost invariant with the increasing number of bits. But the proposed DPSOA-I-GWO does not encounter such problems, where the PEB decreases with the increasing number of bits.
A notable point is that the proposed DPSOA-I-GWO works well even when the number of bits is small. For example, the PEB of DPSOA-I-GWO is about 0.004 m when $N_B=2$ in scenario B.

\begin{figure}
  \centering
  \includegraphics[width=0.45\textwidth]{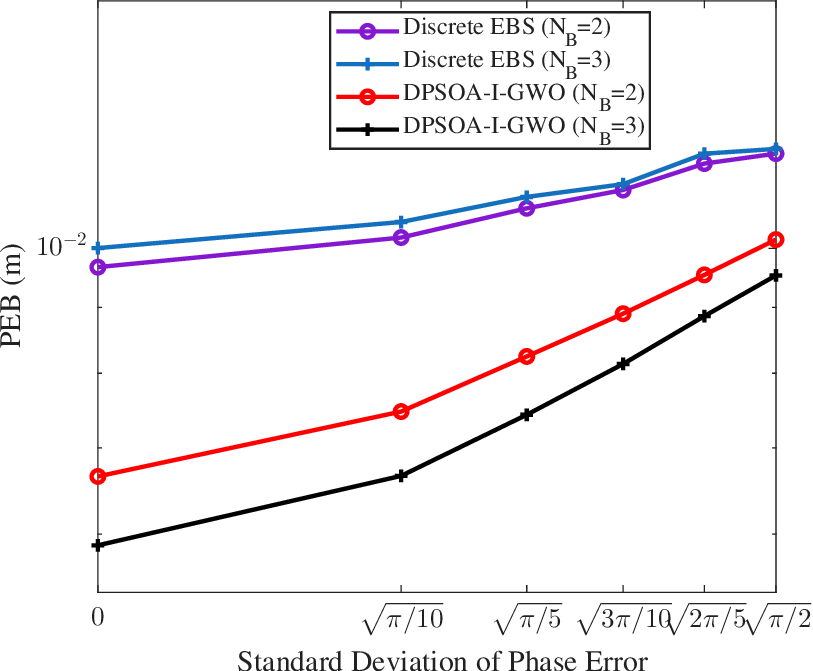}\\
  \caption{PEB versus standard deviation of phase error.}\label{phasenoisefig}
  \vspace{-1\baselineskip}   
\end{figure}

Since the phase error of reflective element, i.e., the error between the actual phase shift and the designed phase shift, is common in practical engineering applications, the influence of phase error on the proposed DPSOA-I-GWO is simulated. The results are shown in Fig.~\ref{phasenoisefig}, which plots the PEB versus the standard deviation of phase error. The phase error is assumed to follow a normal distribution. As expected, the PEB of DPSOA-I-GWO increases significantly as the standard deviation of phase error becomes larger. However, DPSOA-I-GWO always outperforms the discrete EBS. Even when the standard deviation of phase error reaches $\sqrt{\frac{\pi}{2}}$, the PEB of DPSOA-I-GWO is still lower than 0.01 m.

\begin{figure}
  \centering
  \includegraphics[width=0.45\textwidth]{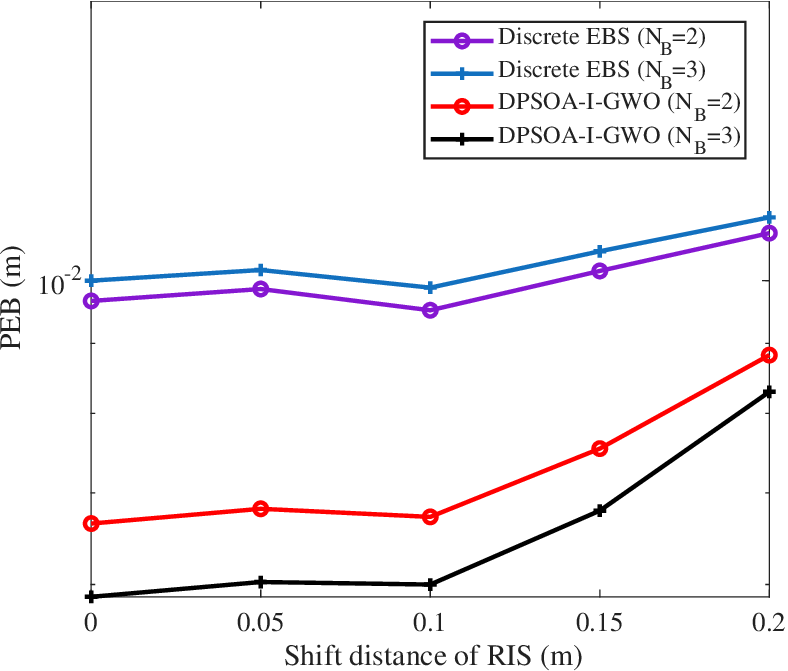}\\
  \caption{PEB versus shift distance of RIS.}\label{phaseshiftRISfig}
  \vspace{-1\baselineskip}   
\end{figure}

Since the RIS geometry may not be perfectly measured in practical engineering applications, the influence of the offset between the real and measured positions of the RIS on the proposed DPSOA-I-GWO is simulated. The results are shown in Fig.~\ref{phaseshiftRISfig}, which plots the PEB versus the shift distance of the RIS. Note that all RISs have the same shift distance, and only the center positions of the RISs are moved in the simulation. This figure shows that when the shift distance is small, its influence on DPSOA-I-GWO is negligible. However, as the shift distance increases, the degradation of PEB becomes significant. For all shift distances considered, DPSOA-I-GWO consistently outperforms the discrete EBS method.

\begin{figure}
  \centering
  \includegraphics[width=0.45\textwidth]{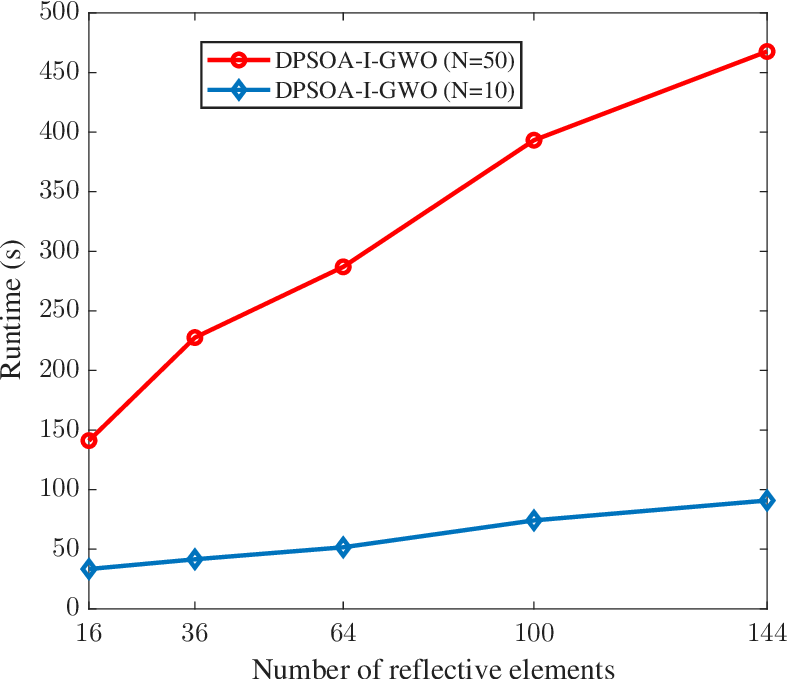}\\
  \caption{Runtime of DPSOA-I-GWO versus number of reflective elements.}\label{rumtimedisfig}
  \vspace{-1\baselineskip}   
\end{figure}

Fig.~\ref{rumtimedisfig} shows the runtime of the proposed DPSOA-I-GWO versus the number of reflective elements. The simulation results are obtained on a host equipped with an Intel i9-14900HX processor. It shows that the runtime of DPSOA-I-GWO is acceptable only when both the number of reflective elements and the number of measurements are at low region.

\section{Conclusion}
In this paper, we have investigated a 3D mmWave positioning system enhanced by multiple RISs. Through a sequential measurement framework exploiting VLoS paths, we have derived the CRLB and PEB to quantify the system's theoretical limits. For continuous RIS implementations, we have proposed CPSOA-RM, a Riemannian manifold-based algorithm that optimizes phase shifts to minimize PEB. Correspondingly, for discrete RIS configurations, we have proposed DPSOA-I-GWO, an efficient heuristic optimization method. Simulation results have demonstrated that both algorithms significantly outperform the existing approaches in terms of PEB. Notably, the results prove that sub-centimeter positioning accuracy is achievable with just two RISs when using the proposed optimization algorithms.

\appendices
\vspace{-0.5cm}
\section{Derivation of Eq.~(\ref{deltadRIS})}
Eq.~(\ref{deltadRIS}) is derived using the near-filed approximation, namely the Fresnel approximation \cite{pan2023ris}, which corresponds to the second-order Taylor expansion. The specific derivation is shown as follows.

\begin{align}
&\Delta d_{I_{i}U,\bar{l}}=d_{I_{i}U,\bar{l}}-d_{I_{i}U}=\|\mathbf{p}_{U}-\mathbf{p}_{I_{i},\bar{l}}\|-d_{I_{i}U}\\\nonumber
&=\left( d_{I_iU}^2 + \Delta x^{2}_{I_i, \bar{l}} + \Delta y^{2}_{I_i,\bar{l}}- 2d_{I_iU}\Phi_{I_iU}\Delta x_{I_i,\bar{l}} \right. \\\nonumber
&~~~~~\left. - 2d_{I_iU}\Psi_{I_iU}\Delta y_{I_i,\bar{l}} \right)^{\frac{1}{2}}-d_{I_{i}U} \\\nonumber
&=d_{I_iU}\left(1+\frac{\Delta x^{2}_{I_i, \bar{l}} + \Delta y^{2}_{I_i,\bar{l}}}{d^2_{I_iU}}\right.    \\\nonumber
&~~~~~~~~~~~~\left.-\frac{2\Phi_{I_iU}\Delta x_{I_i,\bar{l}}+2\Psi_{I_iU}\Delta y_{I_i,\bar{l}}}{d_{I_iU}}\right)^{\frac{1}{2}}-d_{I_{i}U}\\\nonumber
&\stackrel{a}{\approx} \frac{\Delta x^{2}_{I_i, \bar{l}} + \Delta y^{2}_{I_i,\bar{l}}}{2d_{I_iU}}-\Phi_{I_iU}\Delta x_{I_i,\bar{l}}-\Psi_{I_iU}\Delta y_{I_i,\bar{l}}\\\nonumber
&~~~-\frac{\left(\Delta x^{2}_{I_i, \bar{l}} + \Delta y^{2}_{I_i,\bar{l}}\right)^2}{8d^3_{I_iU}}\\\nonumber
&~~~-\frac{d_{I_iU}}{2}\left(\Phi_{I_iU}\frac{\Delta x_{I_i,\bar{l}}}{d_{I_iU}}+\Psi_{I_iU}\frac{\Delta y_{I_i,\bar{l}}}{d_{I_iU}}\right)^2\\\nonumber
&~~~-\frac{\left(\Delta x^{2}_{I_i, \bar{l}}+\Delta y^{2}_{I_i,\bar{l}}\right)}{2d^2_{I_iU}}\left(\Phi_{I_iU}\Delta x_{I_i,\bar{l}}+\Psi_{I_iU}\Delta y_{I_i,\bar{l}}\right)\\\nonumber
&\stackrel{b}{\approx} -\Phi_{I_{i}U}\Delta x_{I_{i},\bar{l}}-\Psi_{I_{i}U}\Delta y_{I_{i},\bar{l}}
+\left(1-\Phi^{2}_{I_{i}U}\right)\frac{\Delta x^2_{I_{i},\bar{l}}}{2d_{I_{i}U}} \\\nonumber
&~~~+\left(1-\Psi^{2}_{I_{i}U}\right)\frac{\Delta y^2_{I_{i},\bar{l}}}{2d_{I_{i}U}}-\Phi_{I_{i}U}\Psi_{I_{i}U}\frac{\Delta x_{I_{i},\bar{l}}\Delta y_{I_{i},\bar{l}}}{d_{I_{i}U}},
\end{align}

where $\mathbf{p}_{I_{i},\bar{l}}$ is the position of the $\bar{l}$-th reflective element of the $i$-th RIS. The approximation (a) is obtained by using the second-order Taylor formula approximation, i.e.,
$\sqrt{1+x}\approx 1+\frac{1}{2}x-\frac{1}{8}x^2$.  Since the distances between UE and  RIS is much larger than the size of RIS reflection element,  the approximation (b) is obtained by neglecting the high-orders of
$\frac{\Delta x_{I_i,\bar{l}}}{d_{I_iU}}$ and $\frac{\Delta y_{I_i,\bar{l}}}{d_{I_iU}}$.

\ifCLASSOPTIONcaptionsoff
  \newpage
\fi
\small
\bibliography{IEEEfull,cite}
\bibliographystyle{IEEEtran}
\end{document}